\documentclass[preprint2]{aastex}

\usepackage{amsmath}
\usepackage{color}

\shorttitle{Cyg OB 7 Cloud Complex}
\shortauthors{Dobashi et al.}

\begin{document}


\title{COLLIDING FILAMENTS AND A MASSIVE DENSE CORE IN THE CYGNUS OB 7 MOLECULAR CLOUD}

\author{
K{\scriptsize AZUHITO} D{\scriptsize OBASHI}\altaffilmark{1}, 
T{\scriptsize OMOAKI} M{\scriptsize ATSUMOTO}\altaffilmark{2},
T{\scriptsize OMOMI} S{\scriptsize HIMOIKURA} \altaffilmark{1}, \\
H{\scriptsize IRO} S{\scriptsize AITO}\altaffilmark{3}, 
K{\scriptsize O} A{\scriptsize KISATO}\altaffilmark{1},
K{\scriptsize ENJIRO} O{\scriptsize HASHI}\altaffilmark{1},
\and
K{\scriptsize EISUKE} N{\scriptsize AKAGOMI}\altaffilmark{1}
}

\affil{\scriptsize{\rm $^1$ dobashi@u-gakugei.ac.jp}}
\altaffiltext{1}{Department of Astronomy and Earth Sciences, Tokyo Gakugei University, Koganei, Tokyo  184-8501, Japan}
\altaffiltext{2}{Department of Humanity and Environment, Hosei University, Fujimi, Chiyoda-ku, Tokyo 102-8160, Japan}
\altaffiltext{3}{National Astronomical Observatory of Japan, Mitaka, Tokyo 181-8588, Japan}



\begin{abstract}

We report results of molecular line observations carried out toward a massive dense core 
in the Cyg OB 7 molecular cloud. The core has an extraordinarily large mass
($\sim1.1 \times 10^4$ $M_\odot$) and size ($\sim2 \times 5$ pc$^2$), but there is no 
massive young star forming therein. We observed this core in various molecular lines such 
as C$^{18}$O($J=1-0$) using the 45m telescope at Nobeyama Radio Observatory. 
We find that the core has an elongated morphology consisting of several filaments and core-like structures.
The filaments are massive ($10^2-10^3$ $M_\odot$), and they are apparently colliding against each other.
Some candidates of YSOs are distributed around their intersection, suggesting that the collisions
of the filaments may have influenced on their formation.
To understand the formation and evolution of such colliding filaments, we performed numerical
simulations using the adaptive mesh refinement (AMR) technique adopting the observed core 
parameters (e.g., the mass and size) as the initial conditions. Results indicate that the filaments 
are formed as seen in other earlier simulations for small cores in literature, but we could not 
reproduce the collisions of the filaments simply by assuming the large initial mass and size. 
We find that the collisions of the filaments occur only when there is a large velocity 
gradient in the initial core in a sense to compress it. We suggest that the observed core 
was actually compressed by an external effect, e.g., shocks of nearby supernova remnants 
including HB21 which has been suggested to be interacting with the Cyg OB 7 molecular cloud.

\end{abstract}



\section{INTRODUCTION}
The Cyg OB 7 molecular cloud is a giant molecular cloud (GMC)
located at a distance of 800 pc \citep{Humphreys1978} in the direction of the Cygnus 
region \citep[e.g.,][]{Falgarone1987}.
The cloud has an apparent size of $\sim 4^\circ  \times 7^\circ$
centered at $\ell \sim92^\circ$ and $b \sim 4^\circ$,
and has a total molecular mass of $\sim1\times 10^5$ $M_\odot$
\citep[][]{Dobashi1994,Dobashi1996}.

\begin{figure*}
\begin{center}
\includegraphics[scale=1.0]{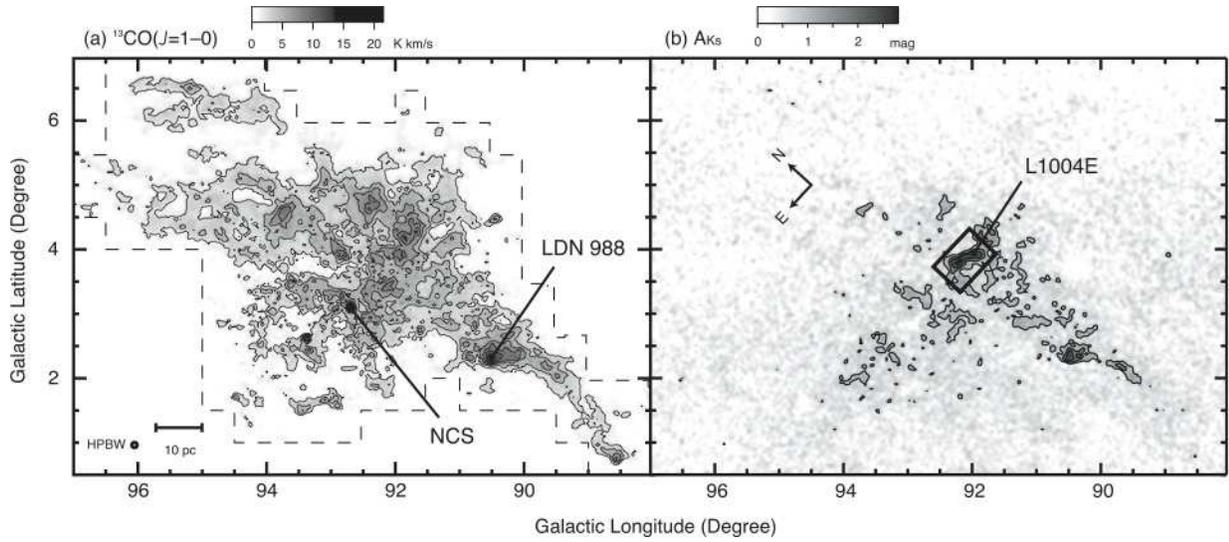}
\caption{
(a) Distribution of the  $^{13}$CO$(J=1-0)$ integrated intensity
of the Cyg OB 7 molecular cloud. Contours start from 2 K km s$^{-1}$ with a 
2 K km s$^{-1}$ step. The data were obtained by the 4m telescope (HPBW$=2\farcm7$) 
at the Nagoya university (Dobashi et al. 2014, in preparation). 
(b) Extinction map of $A_{K_{\rm S}}$ generated using the 2MASS PSC
 \citep{Dobashi2011,Dobashi2013}.  
The lowest contour and the contour interval are $A_{K_{\rm S}}=0.9$ mag and 0.6 mag, 
respectively. The box labeled L1004E indicates the region observed along the 
equatorial coordinates (J2000) with the 45m telescope.
Directions of the north and east on the sky are indicated by the arrows.
\label{fig:location}}
\end{center}
\end{figure*}

Figure \ref{fig:location}(a) shows the entire extents of the cloud.
There are two well-known star forming sites in the cloud.
One is the region known as LDN 988 \citep{Lynds1962} where a number of pre-main 
sequence stars are forming \citep[e.g.,][]{Herbig2006, Allen2008},
and the other is a massive dense core known as the Northern Coal Sack (NCS)
producing a massive Class 0 object  \citep{Bernard1999}.
Except for these two regions, star formation is much less active in this cloud
compared to other GMCs with a similar mass \citep[e.g., Orion A,][]{Nagahama1998},
and there is no extended H{\scriptsize{II}} regions associated with the cloud.
The Cyg OB 7 molecular cloud is also characterized by its low temperature ($\sim 10$ K) and
a large turbulence \citep[$\Delta V \sim 4$ km s$^{-1}$ in $^{13}$CO,][]{Dobashi1994}, 
and has been regarded as a GMC in an early stage of cloud evolution prior to an active 
massive star formation.

In the all-sky visual (VIS) and near-infrared (NIR) extinction maps 
derived from the Digitized Sky Survey \citep[DSS,][]{Dobashi2005} and 
the 2 Micron All Sky Survey \citep[2MASS,][]{Dobashi2011,Dobashi2013},
we found a massive dense core in the northern part of the 
Cyg OB 7 molecular cloud. In the DSS-based extinction map, the core 
is cataloged as TGUH 541P1 by \citet{Dobashi2005} 
which corresponds to the eastern end of the dark nebula 
LDN 1004 \citep{Lynds1962}. In the optically thinner 2MASS-based extinction map,
it splits into several smaller condensations such as 
the one numbered No.2996 in the catalog compiled by \citet{Dobashi2011}. 
Hereafter, we shall refer to this core as L1004E in this paper.

In Figure \ref{fig:location}(b), we show the location of L1004E
in the $A_{K_{\rm S}}$ map produced by \citet{Dobashi2011}. 
As seen in the figure, the core is very large and is the densest in
the entire Cyg OB 7 cloud with the maximum $A_{K_{\rm S}}$ of $\sim3$ mag.
Based on the $A_{K_{\rm S}}$ map,
we estimate its total mass to be at least  $\sim5 \times 10^3$ $M_\odot$. 
In spite of its huge mass, L1004E is not accompanied by 
any H{\scriptsize{II}} regions or bright infrared sources representing massive 
young stellar objects (YSOs),
indicating that L1004E is a core in 
an initial stage of massive star formation or cluster formation.
L1004E should provide us a precious opportunity to investigate the 
initial conditions of such massive cores, because the initial conditions
can be easily destroyed by the strong stellar wind and UV radiation
from OB stars soon after they are formed in the cores.

The angular resolutions of the 2MASS- and DSS- based extinction maps 
($\sim3'-6'$) that we used to find L1004E are 
not sufficient to resolve the structure of the core. 
We therefore carried out molecular line observations at a high angular 
resolution using the 45m telescope (HPBW$\simeq 15"$ at 115 GHz) at 
Nobeyama Radio Observatory (NRO), which should also provide us information
on the velocity field of the core.

The purpose of the present paper is to report results of the observations with the 
45m telescope. We observed the core in various molecular lines such as C$^{18}$O.
The observational procedures are described in Section \ref{sec:observations}.
Based mainly on the C$^{18}$O data, we analyzed the spatial and velocity
structure of the core, and we also searched for candidates of YSOs
associated with the core using some infrared point source catalogs (PSCs) open to the public. 
We found that the core has a huge total mass of $\sim1.1 \times 10^4$ $M_\odot$
and is consisting of a number of massive filaments and core-like structures
with a mass of $10^2-10^3$ $M_\odot$, which appears similar to the
filamentary structures recently evidenced in other molecular clouds \citep[e.g.,][]{Andre2010}.
To our surprise, some of the filaments are apparently colliding against each other,
and some candidates of YSOs are located around the intersections of the filaments
as if they were induced by the collisions of the filaments.
We present these observational results in Section \ref{sec:results}.
In order to understand how the colliding filaments were
formed in the core, we performed numerical simulations of core evolution
assuming a model core with an initial mass and size similar 
to those of L1004E. In section \ref{sec:discussion}, we describe
the method and results of the simulations, and discuss what 
initial conditions are needed for the formation of such colliding 
filaments which should play an important role for massive star 
formation and cluster formation in massive cores.
Conclusions of this paper are summarized in Section \ref{sec:conclusions}.


\section{OBSERVATIONS} \label{sec:observations}

Observations were carried out with the 45m telescope
at NRO.  We observed thirteen molecular lines in total,
i.e., $^{12}$CO$(J=1-0)$, $^{13}$CO$(J=1-0)$, C$^{18}$O$(J=1-0)$,
CS$(J=2-1)$, C$^{34}$S$(J=2-1)$, HCO$^{+}(J=1-0)$, H$^{13}$CO$^{+}(J=1-0)$,
HC$_3$N$(J=5-4)$, CCS$(J_{N}=4_3-3_2)$, 
NH$_{3} (J,K=1,1)$, NH$_{3} (J,K=2,2)$, NH$_{3} (J,K=3,3)$, and NH$_{3} (J,K=4,4)$.
The $^{13}$CO$(J=1-0)$ emission line was observed for 4 days in 2010 January,
and the other lines were observed for 22 days in the period between 2009 February and
2009 May.

Our observations can be divided into two modes. One is the mapping observations
to reveal the entire molecular distributions of L1004E, and the other is one-point
observations made only toward the protostellar candidate IRAS 21025+5221 found in the core.

The mapping observations were carried out with 6 molecular lines
as summarized in Table \ref{tab:mapping_obs}. We used the 25 Beam Array Receiver System (BEARS)
to observe the emission lines of CO and its isotopes. We also used an SIS receiver named S40
to observe the HC$_3$N and CCS lines at 45 GHz as well as a cooled HEMT receiver named H22
to observe the NH$_{3}(J,K)=(1,1)$ line at 24 GHz. Spectrometers were auto-correlators (AC) 
covering a band-width of either 32 MHz or 16 MHz with 1024 channels with
a frequency resolution of 38 kHz or 19 kHz, respectively.
We mapped an area of  $\sim 20' \times 30'$ around the core using the On-The-Fly (OTF)
technique \citep{Sawada2008}, and calibrated the spectral data with the standard chopper-wheel method \citep{Kutner1981}.
Reference positions (i.e., the emission-free OFF positions) were
($\alpha_{\rm J2000}$, $\delta_{\rm J2000}$)$=$
(20$^{\rm h}$38$^{\rm m}$33.9$^{\rm s}$, 51${\arcdeg}$47${\arcmin}$02${\arcsec}$),
(20$^{\rm h}$49$^{\rm m}$34.5$^{\rm s}$, 51${\arcdeg}$48${\arcmin}$00${\arcsec}$),
(20$^{\rm h}$58$^{\rm m}$37.0$^{\rm s}$, 51${\arcdeg}$36${\arcmin}$59${\arcsec}$),
(21$^{\rm h}$04$^{\rm m}$06.5$^{\rm s}$, 52${\arcdeg}$33${\arcmin}$47${\arcsec}$) 
for the $^{12}$CO, C$^{18}$O, NH$_3$, and the other emission lines, respectively.
We used the reduction software package NOSTAR available at NRO 
to subtract a linear baseline from the raw data and to resample 
the spectral data onto $10{\arcsec}$ or $30{\arcsec}$ grid along the equatorial coordinates.
We then applied a correction for the main beam efficiencies which vary in the range $\eta_{\rm mb} =31-84$ \%
depending on the frequencies to the baseline-subtracted data in order to scale them to units of $T_{\rm mb}$.
Further corrections for the side-band ratio were applied for the data obtained by BEARS, because
it is a double-side-band (DSB) receiver.
Velocity channels are resampled onto 0.1 km s$^{-1}$ or 0.2 km s$^{-1}$ velocity grid, 
which resulted in a velocity resolution of $0.14-0.34$ km s$^{-1}$. Noise levels of the resulting data 
are $\Delta T_{\rm mb}=0.08-1.25$ K at these velocity resolutions as summarized in Table \ref{tab:mapping_obs}.

One-point observations toward IRAS 21025+5221 were made with
molecular lines as summarized in Table \ref{tab:point_obs}. 
For these observations, we used waveguide-type dual-polarization sideband-separating 
SIS receivers called T100H/V \citep{Nakajima2008} to observe CS, C$^{34}$S, HCO$^{+}$,
and H$^{13}$CO$^{+}$ lines, and also used the H22 receiver to observe  four NH$_{3}$
lines of $(J,K)=(1,1)-(4,4)$ transitions.
Spectrometers were acousto-optical spectrometers (AOSs) with a bandwidth of 40 MHz 
and a frequency resolution of 37 kHz corresponding to a velocity resolution of $0.11-0.47$ km s$^{-1}$.
The baseline removal was done by using the reduction package NEWSTAR.
The data were scaled up to units of $T_{\rm mb}$ in the same way as for the mapping data.

The system noise temperatures $T_{\rm SYS}$  were in the range $140-430$ K depending on
the receivers including the atmosphere. Pointing accuracy was better than $\sim 10\arcsec$ as was checked by 
observing the SiO maser T-Cep at 43 GHz every 2 hours during the observations.


\section{RESULTS} \label{sec:results}

\subsection{Overall Molecular Distributions \label{sec:distributions}}

Figure \ref{fig:ii_maps} shows the integrated intensity distributions of the
observed emission line listed in Table \ref{tab:mapping_obs}.
As seen in the figure, the C$^{18}$O map in panel (a) reveals 
the filamentary structure of the core.
On the other hand, the $^{13}$CO map in panel (b)
shows rather flat distribution, because the line is heavily saturated.
The $^{12}$CO emission line (not shown) extends all over the mapped area,
and it appears much flatter. 
The other molecular emission lines, ie., HC$_3$N, CSS, and NH$_3$ 
in panels (c)--(e), show clumpy distributions, and they are concentrated 
especially around IRAS 21025+5221. 

In Figure \ref{fig:w3}, we show the 12 $\micron$ image taken by the WISE satellite.
It is noteworthy that the C$^{18}$O map in Figure \ref{fig:ii_maps}(a) is very similar to the dust distribution
revealed by WISE, indicating that the molecular line is a good tracer 
of the total molecular column density $N({\rm H_2})$.  In order to estimate the mass 
of L1004E, we therefore 
derived  $N({\rm H_2})$ at each observed position 
by analyzing the  C$^{18}$O data 
using a standard method assuming the Local Thermodynamic Equilibrium 
\citep[LTE, e.g., see][]{Shimoikura2012}.
For this, we first estimated the excitation temperature of C$^{18}$O 
from the $^{12}$CO spectra at each observed position
assuming that the  $^{12}$CO emission line is optically very thick. 
We then calculated $\tau$(C$^{18}$O) the optical depth of the C$^{18}$O emission line
and its column density $N$(C$^{18}$O), and converted $N$(C$^{18}$O) to 
$N$(H$_{2}$) using an empirical conversion relation found by \citet{Frerking1982}. 
We summarize the method in the Appendix.

The analyses of the C$^{18}$O and $^{12}$CO data infer that 
L1004E has a huge mass of $\sim 1.1\times10^4$ $M_\odot$
within the region mapped in C$^{18}$O (Figure \ref{fig:ii_maps}(a)).
This mass is much higher than other single dense cores found 
in low-mass star forming regions such as Taurus \citep[$1-80$ $M_\odot$,][]{Onishi1998}
or in massive star forming regions found in the vicinity of some H{\scriptsize{II}}
regions \citep[$10^2-10^3$ $M_\odot$,][]{Tachihara2002,Saito2007,Shimoikura2013}
suggesting that L1004E may comprise a number of distinct smaller cores.
Actually, as shown in Section \ref{subset:filaments}, 
L1004E is very likely to consist of several massive filaments having a 
mass of $10^2-10^3$ $M_\odot$. 

It is also noteworthy that the C$^{18}$O column density of L1004E
is very high compared with cores in other star forming regions. 
In fact, we found that the C$^{18}$O emission line which is often
optically thin ($\tau<1$) in molecular clouds has a high optical depth of $\tau =2-3$
at the peak intensity positions in Figure \ref{fig:ii_maps}(a).
Figure \ref{fig:column_density} shows the frequency distributions of $N$(C$^{18}$O) in L1004E
compared with those of HLC 2 and LDN 1551 observed by the same NRO 45m telescope
(the data of these regions are open to the public at the website of NRO, 
http://www.nro.nao.ac.jp/$^{\backsim}$nro45mrt/results/data.html).  
As seen in the figure, L1004E exhibits much
higher $N$(C$^{18}$O) than the others.  On the basis of statistical studies of dense cores in Taurus,
\citet{Onishi1998} showed that all of the cores with column density greater than
$N$(H$_2$)$= 8\times10^{21}$ cm$^{-2}$ are accompanied by YSOs
selected from IRAS sources,
and they suggested that the cores start to form stars immediately as soon as
the column density exceeds the value.
This critical value of $N$(H$_2$) corresponds to 
$N$(C$^{18}$O)$\sim 8 \times 10^{14}$ cm$^{-2}$ (see Equation (\ref{eq:NH2}) in the Appendix).
As shown in the next subsection,
there are only two IRAS sources as promising candidates of protostars
in this massive core, suggesting that the core may be in a stage prior to an active star formation
or may have just initiated forming stars.

L1004E is also characterized by its low temperature.
Within the observed region, the excitation temperature derived from the $^{12}$CO 
emission line varies in the range
$8 \lesssim T_{\rm ex}\lesssim12$ K
with a mean value of $\sim 9$ K, which is 
lower than those in other GMCs with a similar size
\citep[e.g., $10-70$ K,][]{Nagahama1998},
but close to those in smaller dark clouds such as the one in Taurus
\citep[e.g., $\sim 10$ K,][]{Onishi1998}.
In addition, there is a clear tendency that 
$T_{\rm ex}$ is lower in the central part of the core 
than in the outskirts by a few Kelvin.
Similarly, the observed line width measured by applying a single Gaussian fitting 
to the C$^{18}$O emission line varies in the range $1 \lesssim \Delta V({\rm C^{18}O}) \lesssim 3 $ km s$^{-1}$
with a mean value of $\sim$2 km s$^{-1}$ in the region surrounded by
the lowest contour in Figure \ref{fig:ii_maps}(a), and it tends to be smaller
where the C$^{18}$O intensity is higher, suggesting dissipation of turbulence
in the central part of the core.

We summarize the global properties of L1004E found through 
the above $^{12}$CO and C$^{18}$O analyses in Table \ref{tab:core}.

Finally, we show the velocity distributions of molecular gas in a series
of channel maps in Figures \ref{fig:channelmap_13co_0}--\ref{fig:channelmap_c18o_1}.
Molecular emission lines, especially the $^{12}$CO line (not shown), spread over a wide velocity
range of $ -20 \lesssim V_{\rm LSR} \lesssim 10$ km s$^{-1}$, but the other 
emission lines concentrate in a rather limited range around $V_{\rm LSR} \simeq -2$ km s$^{-1}$ 
(see Figure \ref{fig:channelmap_13co_0} displaying the $^{13}$CO distributions). 
Figures \ref{fig:channelmap_13co_1} and \ref{fig:channelmap_c18o_1} show channel maps 
of $^{13}$CO and C$^{18}$O, respectively, produced at a higher velocity resolution of 
0.5 km s$^{-1}$.  These figures show that denser parts of the core exhibits an elongated  
structure, and the core apparently consist of a number of filaments having
slightly different velocities. The filaments are clearer in the C$^{18}$O channel maps 
in Figure \ref{fig:channelmap_c18o_1}. We will attempt to identify the individual filaments and 
will further analyze their stability and interactions in Section \ref{subset:filaments}.

\subsection{Young Stellar Objects \label{subset:protostars}}

In order to search for YSOs in the observed region, 
we selected candidates of YSOs in the IRAS PSC
using the criteria suggested by \citet{Onishi1998}
for a search of YSOs in the Taurus cloud complex:
We selected sources detected at least 3 bands including 25 $\micron$ and 60 $\micron$
satisfying the conditions log$(F12/F25)<0.0$ and log$(F25/F60)<0.3$
where $F12$ means a flux density at 12 $\micron$, and so on. 
As a result, we found only 2 IRAS sources in the observed 
region. They are IRAS 21005 + 5217 and 21025 + 5221 whose properties are
listed in Table \ref{tab:iras}. 
Both of the IRAS sources have a cold FIR spectra
typical of YSOs \citep[e.g.,][]{Fukui1989}, having a total far-infrared (FIR) luminosity
$L_{\rm IRAS}\simeq 22$ and $\sim 36$ $L_\odot$.
Note that these are the bolometric luminosities detected only in the four IRAS bands
\citep[including the correction for wavelengths longer than100 $\micron$,][]{Myers1987},
and the flux in the wavelengths shorter than 12 $\micron$ is not included.

We further searched for YSOs in the WISE PSC 
using the selection criteria suggested by \citet{Koenig2012},
which is to select candidates of Class I and Class II sources \citep[][]{Lada1987,Greene1994}
from the WISE PSC.
The criteria require detection in the WISE 3.4 $\micron$, 4.6 $\micron$, and 12 $\micron$ bands with
rather complex limitations in magnitudes and colors in the 3 bands to exclude
non-YSOs objects such as galaxies, active galactic nuclei (AGNs), unresolved knots of shock emission
due to outflows, and so on \citep[for details, see the Appendix of][]{Koenig2012}.
As a result, we found a score of candidates for Class I or Class II sources
in the observed region, but they are mostly located outside of the denser parts of the core, 
suggesting that many of them are the sources either in the foreground or background,
not having physical association with the core.

We should note, however, that there may be more faint YSOs in the WISE PSC, 
which were excluded by the tight selection criteria of \citet{Koenig2012}.
As they state in their paper, it is difficult in general to establish definite criteria to select YSOs perfectly.
Their criteria have a strong restriction especially on the magnitudes of the 
sources in order to exclude AGNs, i.e., the apparent magnitudes at 3.4 $\micron$
and 4.6 $\micron$ have to be brighter than 14.0 and 13.5 mag, respectively
(see their Section A.1). If we disregard these restrictions to exclude AGNs, we would find more 
candidates of faint YSOs.  Actually, we found a score of such faint 
sources lying along the filaments of the core,
which are likely to be YSOs forming therein.
In the original WISE PSC, however, it seems that there are a certain fraction of false detections
probably caused by the algorithm to extract point sources from the WISE images.
In order to avoid such false detections, we checked by eyes the locations of the point sources
on the WISE images, and selected only those having an apparent counterpart at least in one of the
four band images of WISE (i.e., 3.4, 4.6, 12, and 22 $\micron$).
As a result, we found  83 candidates of YSOs (i.e., 50 Class I sources and 33 Class II sources)
within the area mapped in C$^{18}$O including the above faint sources without the restrictions to
exclude AGNs. The 2 IRAS sources
are also included in these numbers because they have a counterpart in the selected WISE sources.
We show locations of the these sources in Figure \ref{fig:ysos}, and we shall regard them as YSOs forming in L1004E.

Here we attempt to estimate the star formation efficiency (SFE) of L1004E.
A certain fraction of the 83 sources should be AGNs or YSOs unrelated to the core, 
but for simplicity, we shall assume that all of them are YSOs forming in the core.
We also assume that their average mass is $\sim1$ $M_\odot$ 
which is the mean value of stars following the Salpeter's mass function 
\citep[$dN/dm \propto m^{-2.35}$, ][]{Salpeter1955}
over the range $0.4-10$ $M_\odot$. 
These assumptions yield an estimate of the SFE of only $\lesssim1$ \%,
because the total molecular mass of the core is
$1.1\times 10^4$ $M_\odot$ (see Table \ref{tab:core}). 
This value of SFE is much lower compared with other massive cores forming
clusters \citep[e.g., SFE$\sim30$ \% on the average, ][]{Shimoikura2013}.
Note that this is the maximum estimate for the SFE, because a certain fraction
of the WISE sources should be unrelated to L1004E.
In an extreme case, if we take that only the 2 IRAS sources and
several bright WISE sources which are located where the C$^{18}$O emission is strong
are promising YSOs forming in L1004E,
a very low SFE of only $\sim0.1$ \% would inferred, which should be the minimum estimate of the SFE
in L1004E. 


We found that the 2 IRAS sources selected as candidates of YSOs
have a counterpart not only in the WISE PSC
but also in the 2MASS PSC, 
as summarized in Table \ref{tab:counterparts}. 
IRAS 21025 + 5221 is probably a younger YSO than IRAS 21005 + 5217,
because it is not detected in the $J$ and $H$ bands of 2MASS,
while IRAS 21005 + 5217 is not detected in the IRAS 100 $\micron$ band.
In order to better access the stellar properties of these sources
such as the ages, masses, and luminosities, 
we employed a recent stellar model developed 
by Robitaille et al. \citep{Robitaille2007,Robitaille2008}.
The model was first developed by \citet{Robitaille2006}, 
and it comprises several parameters for the central star, circumstellar disk, and envelope,
to calculate the spectral energy distributions (SEDs) for a wide range of
wavelengths that can be compared with the observations. 
Tools for the model fitting are available on their website (http://caravan.astro.wisc.edu/)
which provide us 10,000 sets of the model parameters to fit observational data 
sorted according to the resulting $\chi^2$.
We utilized their tools on the web to fit the SEDs of the 2 IRAS sources 
using the observed parameters listed in Tables \ref{tab:iras} and \ref{tab:counterparts}.
Resulting SEDs of the model best fitting the data are shown in
Figure \ref{fig:sed}, and 
some of the best model parameters with the minimum $\chi^2$ are summarized 
in Table \ref{tab:modelpara}. The $A_V$ values in the first column
of the table are the fitted total extinction along the line-of-sight to the stellar envelope, 
which is consistent with what we would expected from the observed column 
density of C$^{18}$O. The other parameters (age, mass, and luminosity)
are the best fitting parameters for the central star. Looking into the best 10 sets of
the model parameters with the smallest $\chi^2$, these parameters summarized 
in the table seem to be well 
determined with small variations in the case of IRAS 21005+5217,
while the parameters are rather ambiguous in the case of IRAS 21025 + 5221 
varying in the range  $1\times 10^3 - 8 \times 10^4$ yr, $0.2-5$ $M_\odot$, and $12-65$ $L_\odot$,
which is mostly due to the lack of the data points in the $J$ and $H$ bands
(there are only the upper limits for these data points). 
In any case, IRAS 21025 + 5221 is obviously younger than
the other source.

\subsection{Dense Gas  Around IRAS 21025+5221  \label{subset:iras21025}}

We found that IRAS 21025+5221 
is accompanied by a small and well-defined molecular condensation which should be the direct parental core
of the IRAS source, while the other IRAS source is not accompanied by such a very evident condensation.
Figure \ref{fig:IImaps_IRAS21025} displays the close up view
of the condensation traced by some molecular emission lines,
and Figure \ref{fig:Sample_spectra} shows spectra of various molecular 
lines observed toward the IRAS source.

As seen in Figure \ref{fig:IImaps_IRAS21025}, the condensation is 
evident especially in HC$_3$N, CCS, and NH$_3$.
Ratio of the NH$_3(J,K)=(1,1)$ and $(2,2)$ emission lines
infers the gas temperature of the condensation of $\sim12$ K
\citep[e.g., ][]{Ho1983, Danby1988}, which is close to the excitation 
temperature $T_{\rm ex}\simeq 10$ K derived from the $^{12}$CO emission line.

As shown in Section \ref{subset:filaments},
we estimate the total mass, radius, and mean
molecule density of the condensation
to be $M_{\rm LTE}\simeq170$ $M_\odot$,   $R\simeq0.33$ pc, and $n({\rm H_2})\simeq1.9\times 10^4$ cm$^{-3}$ (see No.1 in Table \ref{tab:filaments}), respectively,
based on the C$^{18}$O data
in the same way as described in Section \ref{sec:distributions}.
Note that the condensation is accompanied by 
IRAS 21025 + 5221 and many other 
faint YSOs selected from the WISE PSC,
indicating that the condensation is probably a young dense core that 
has just initiated to form a star cluster.

The idea that the condensation is very young can be supported also by
the CCS and HC$_3$N data. It is known that these molecules
are abundant in an early stage of chemical evolution of dense cores ($\sim 10^5$ yr), 
and they rapidly disappear along with the core evolution \citep[$\gtrsim 10^6$ yr,][]{Suzuki1992,Hirota2009}.
Based on the CCS and HC$_3$N spectra shown in Figure \ref{fig:Sample_spectra},
we calculated the fractional abundances of these molecules, 
i.e., $f(X)=N(X)/N({\rm H}_2)$ where $X=$CCS or HC$_3$N,
using the same method as \citet[][see their Section 3.2]{Shimoikura2012},
and we found $\log f({\rm CCS})=-9.09 $ and $\log f({\rm HC_3N})=-9.45$.
We compare the results with those measured in dense cores in other star 
forming regions as well as with a model calculation performed by \citet{Suzuki1992} 
in Figure \ref{fig:abundances} which is quoted from
Figure 9 of  \citet{Shimoikura2012} who studied starless cores in the Polaris cirrus
(see their Table 9 for data points of the other cores).
There is an apparent tendency in the figure that 
data points for dense cores not associated with YSOs 
(represented by the open circles)
are located in the upper-right side of the diagram having
higher $f({\rm CCS})$ and $f({\rm HC_3N})$, while
those associated with YSOs (filled circles) are
more widely distributed in rather bottom-left side.
It is interesting to note that the condensation associated with 
IRAS 21025 + 5221 (the open square labeled L1004E)
is located between the two distributions, suggesting that the condensation 
is a young core which has just started star formation.

Finally, we should note that there are some noticeable features in a series of spectra 
shown in Figure \ref{fig:Sample_spectra}. Most of the spectra have a peak 
radial velocity of $V_{\rm LSR}\sim -2$ km s$^{-1}$ which should represent the systemic
velocity of gas around IRAS 21025+5221. However, the HCO$^+$ and CS emission lines
exhibit a double-peaked profile with a greater blue-shifted velocity component.
Such a profile is often seen around protostars in optically thick emission lines like HCO$^+$, 
and is considered as a sign of collapsing motion of dense cores \citep[e.g.,][]{Zhou1993}.
It is also noteworthy that the $^{12}$CO emission exhibits a very broad 
line width extending over the velocity range $-20<V_{\rm LSR}<10 $ km s$^{-1}$. 
This could be due to a molecular outflow possibly associated with IRAS 21025+5221,
although it is difficult to confirm its red and blue shifted wings
because the $^{12}$CO line around the IRAS source is heavily contaminated 
by the emission from the diffuse ambient gas over a wide velocity range (especially
at $-20<V_{\rm LSR}<-10 $ km s$^{-1}$).
The wing-like features are also seen in the spectra of HCO$^{+}$ and CS which should be 
more free from such contamination, but we have only limited data for these lines.
Observations of these molecular lines at a higher angular resolution using interferometers
such as SMA and BIMA would be needed to confirm and resolve the outflow as well as 
the collapsing motion of the condensation.

\subsection{Giant Filaments \label{subset:filaments}}

\subsubsection{Identification of the Filaments and Their Gravitational Stability \label{subsubsect:stability}}

As seen in the C$^{18}$O total intensity map in Figure \ref{fig:ii_maps}(a),
L1004E has an elongated morphology consisting of several filaments and core-like structures.
In order to characterize the filaments, we attempted to identify the individual
filaments based on the C$^{18}$O data. We first tried to find a systematic way
to define the filaments,
but we found that it is very difficult to find a numerical definition to separate the filaments reliably.
After some trials, we finally decided to pick up relatively large filaments by eye inspection
in an iso-temperature map of C$^{18}$O in 3 dimension (Figure \ref{fig:filaments}). 
We then determined the velocity range where the identified filaments are detected, 
and produced a C$^{18}$O intensity map integrated over the velocity range
for each of the filaments to measure their physical properties such as the mass, size, and line width.
Although our method to define the filaments is tentative,  we will analyze the filaments
identified in this manner, because our main purpose is not to compile a complete list of the filaments
but to investigate their typical properties, gravitational stabilities, and velocity structures.

In total, we picked up 12 filaments as labeled in Figure \ref{fig:filaments}.
The condensation associated with IRAS 21025 + 5221 shown in 
Section \ref{subset:iras21025} corresponds to the one labeled No.1.
Some of the filaments including the condensation have a round shape, 
and it may be more appropriate to call them ``cores" or ``subcores" rather 
than ``filaments".
In order to see how much the filaments are round or elongated,
we fitted the C$^{18}$O intensity distribution of the individual filaments
by an elliptic 2 dimensional Gaussian function, and derived their major and minor 
radii ($R_{\rm maj}$ and $R_{\rm min}$) to determine their ellipticities $\alpha$ ($=R_{\rm min}/R_{\rm maj}$)
which can be a measure of their elongated shapes (e.g., $\alpha=1$ is for a spherical shape).
In this subsection, we call filaments with $\alpha \le 0.5$ ``filaments"
and call the others with with $\alpha > 0.5$ ``subcores". Four of the 12 filaments we identified
(Nos. 1, 2, 8, and 9 in Figure \ref{fig:filaments}) can be classified to subcores.

Other than the ellipticities, we
measured some properties of the filaments and subcores which are summarized in Table \ref{tab:filaments}. 
In the table, the parameter $V_{\rm range}$ is the velocity
range that we used to define the filaments and subcores, and $\alpha_{\rm J2000}$ and $\delta_{\rm J2000}$
are their peak positions in the C$^{18}$O intensity maps integrated over $V_{\rm range}$.
$N$(H$_2$) and $\Delta V$ are the molecular column density and the line width (defined at FWHM)
measured at the peak positions, respectively,
and $M_{\rm LTE}$ is the mass derived from the C$^{18}$O intensity 
in the same way as described in Section \ref{sec:distributions}. We defined the surface area
of the filaments and subcores $S$ at the half maximum of the peak C$^{18}$O intensity
integrated over $V_{\rm range}$, 
and defined their radii as $R=\sqrt{S/\pi}$. We also derived
C$^{18}$O spectra averaged over $S$, and fitted them 
by Gaussian functions to measure the mean line width $\Delta {\overline V}$. 
In order to check the gravitational stability of the filaments and subcores, we also 
calculated their Virial mass in a standard way \citep[e.g., ][]{Kawamura1998} as
\begin{equation}
\label{eq:Mvir}
{M_{\rm vir}} = 209\left( {\frac{R}{{\rm pc}}} \right){\left( {\frac{{\Delta {\overline V}}}{{\rm km~{s^{ - 1}}}}} \right)^2}~~M_\odot~~.
\end{equation} 

As summarized in the table, we found that the identified filaments and subcores have  a mass, line width, 
and column density of $M_{\rm LTE}=107-793$ $M_\odot$, $\Delta {\overline V}=1.0-2.9$ km s$^{-1}$, and 
$N({\rm H_2})=1.9-4.5\times10^{22}$ cm$^{-2}$ with a mean value of 
341 $M_\odot$, 1.7 km s$^{-1}$, $2.5\times10^{22}$ cm$^{-2}$, respectively.
These quantities are comparable to those of cores forming massive
stars \citep[e.g.,][]{Tachihara2002,Saito2007} or star clusters \citep[e.g.,][]{Higuchi2010,Shimoikura2013}.

We also found that the masses of the filaments and subcores
$M_{\rm LTE}$ are roughly consistent with 
$M_{\rm vir}$ as shown in Figure \ref{fig:Mvir}(a), indicating that they are 
likely to be in the Virial equilibrium. However,  dispersion in the $M_{\rm vir}$ vs. $M_{\rm LTE}$
diagram is large, probably because $M_{\rm vir}$ in Equation (\ref{eq:Mvir}) is for 
an ideal sphere while many of the filaments and subcores
have an elongated shape. Actually, the difference
between $M_{\rm LTE}$ and $M_{\rm vir}$ tends to be larger for the
filaments and subcores
with lower ellipticities, e.g., $M_{\rm LTE}$ is a few times higher than $M_{\rm vir}$ for
many of the filaments with $\alpha \le 0.5$.

In order to better investigate the gravitational stability of highly elongated filaments, 
we employed a simple model of isothermal cylindrical gas with infinite length
\citep[e.g.,][]{Stodolkiewicz1963,Ostriker1964,Inutsuka1992}. 
The cylindrical gas can be gravitationally stable when the mass density $\rho$ as 
a function of radius $r$ from its axis follows the equation
\begin{equation}
\label{eq:cylindrical}
\rho (r) = {\rho _c}{\left[ {1 + {{\left( {\frac{r}{{{H_0}}}} \right)}^2}} \right]^{ - 2}}
\end{equation} 
where $H_0$ is an effective radius written as
\begin{equation}
\label{eq:H0}
{H_0} = \sqrt {\frac{2}{{\pi G{\rho _c}}}} c_s~,
\end{equation}
and $G$, $c_s$, and ${\rho _c}$ are the gravitational constant, the
speed of sound, and the density at $r=0$, respectively. 
Integration of Equation (\ref{eq:cylindrical}) 
from $r=0$ to infinity yields $M_{\rm cr}$
the mass per unit length of the gravitationally stable filaments,
which can be expressed as
\begin{equation}
\label{eq:Mcr}
M_{\rm cr} =\pi H_0^2 \rho_c= \frac{2c_s^2}{G} ~.
\end{equation}

If we assume that $c_s$ is equal to the observed line width $\Delta V$
because the internal motion of the observed filaments is 
apparently dominated by turbulence rather than thermal motion,
$M_{\rm cr}$ can be written in the following useful form,
\begin{equation}
\label{eq:Mcr2}
M_{\rm cr} = 465{\left( {\frac{{\Delta V}}{{\rm km~{s^{ - 1}}}}} \right)^2} ~~M_\odot~{\rm pc^{-1}}~~.
\end{equation}

Using the above equation, we calculated $M_{\rm cr}$ of the filaments by
inserting $\Delta V$ in Table \ref{tab:filaments}, assuming that we are 
observing the filaments orthogonally to their elongation.
We also calculated the observed gas mass per unit length
of the filaments as $M_{\rm LTE}/L$ where $L$ is the lengths of
the filaments estimated as $L=2R/\sqrt{\alpha}$, which can be compared
with $M_{\rm cr}$. 
We summarize these values in the last columns of Table \ref{tab:filaments},
and compare $M_{\rm cr}$ and $M_{\rm LTE}/L$ in Figure \ref{fig:Mvir}(b).
As seen in the figure,  $M_{\rm cr}$ and $M_{\rm LTE}/L$ do not match
for the subcores with high ellipticities ($\alpha > 0.5$ shown by open circles),
because $M_{\rm cr}$  is almost meaningless for the subcores.
About half of the elongated filaments ($\alpha \le 0.5$, filled circles) 
have $M_{\rm LTE}/L$ close to $M_{\rm cr}$, indicating
that they can be gravitationally stable. The other half of the elongated filaments
(such as No.12) have $M_{\rm cr}$ a few times higher than $M_{\rm LTE}/L$, suggesting that they
might be gravitationally unbound being dispersed into the interstellar medium. 

However, we should note that the C$^{18}$O spectra of such filaments
with $M_{\rm cr}> M_{\rm LTE}/L$
are often contaminated by faint emission from other 
filaments at different velocities. It is difficult to separate them reliably to measure $\Delta V$ precisely
at the present sensitivity of our  C$^{18}$O data.
Such contamination can easily cause an overestimation of $M_{\rm cr}$ 
by a factor of a few. In fact, the difference between $M_{\rm cr}$ and $M_{\rm LTE}/L$ is larger for
filaments with larger $\Delta V$. 
More sensitive C$^{18}$O dataset with much better signal-to-noise ratio is needed to 
verify the gravitational stability of the filaments, 
especially for those appearing gravitationally unbound.

\subsubsection{Velocity Distributions and Collisions of the Filaments \label{subsubsect:collision}}

As can be seen in the C$^{18}$O channel map in Figure \ref{fig:channelmap_c18o_1},
the filaments and subcores often have slightly different radial velocities. 
It is noteworthy that some of them show an apparent
anti-correlation to each other in the channel map, i.e.,
ridge of a filament corresponds to valley or hole of another filament.
A typical example of the anti-correlations is shown
in Figure \ref{fig:collision}.
Filaments labeled 4 and 5 (Table \ref{tab:filaments})
in panel (b) of the figure correspond to a hole
between subcore 2 and filament 6 in panel (a).
Similarly, another filament labeled 7 in panel (b) corresponds
to a hole in fainter C$^{18}$O intensity distribution in panel (a).
We suggest that some of these anti-correlations represent collisions
between the filaments and subcores.

In oder to find an evidence for such collisions, 
we made position-velocity (PV) diagrams along the cuts labeled 
A--B, C--D, and E--F in panel (a) of Figure \ref{fig:collision}.
Resulting diagrams are displayed in panels (c)--(e) of the figure.
As seen in panel (c), the filaments 4 and 5 have 
a similar velocity, and the filament 6 is located in between
having slightly shifted peak velocity by $\sim$0.5 km s$^{-1}$, and
the three filaments appear physically connected to each other 
with fainter C$^{18}$O emission. Another PV diagram in panel (d) 
for the cut C--D shows a similar feature. These features seen in the PV diagrams 
can be naturally accounted for if we assume that the filaments 4 and 5 used to be one 
continuous filament and the subcore 2 and filament 6 used to be another continuous
filament, and they collided and passed through each other.
The filaments 4 and 5 in panel (b)
are shifted by $\sim 0.5'$ ( $=$ 0.12 pc) 
from the corresponding hole in panel (a). If we take that 
the relative velocity of the filaments to be 0.5--1 km s$^{-1}$
as seen in panels (c) and (d), their collision may have 
occurred 0.1 -- 0.2 Myr ago.

As described in Section \ref{sec:discussion}, we have performed numerical simulations
to investigate the formation and evolution of the filaments in a very massive core
having the same total mass as L1004E ($\sim1\times10^4$ $M_\odot$). 
The simulations can reproduce filaments inside of the model core, and some of them 
collide against each other to form stars. It is noteworthy that PV diagrams 
taken around such colliding filaments in the simulations (e.g., see Figure \ref{fig:simulation_color}(c)) 
often appear very similar to those actually observed as displayed in Figure \ref{fig:collision}(c) and (d), 
strongly suggesting that the anti-correlations seen in the C$^{18}$O channel map (Figure \ref{fig:channelmap_c18o_1}) represent collisions of the filaments.

In the case of the anti-correlation seen around the filament 7 in Figure \ref{fig:collision}
(a) and (b), however,
there is no abrupt jump in velocity in the PV diagram (see panel (c)).
Although such a PV map can also be seen in our simulations for a filament drifting
in the shear or for distinct filaments colliding with a small relative velocity, 
it is difficult to judge whether the anti-correlations in the channel maps are due to 
collisions of the filaments or merely trace velocity gradients along a single filament 
without collisions.

We will further investigate the velocity structure of the filaments and subcores
around the 2 IRAS sources. As shown in Section \ref{subset:iras21025},
IRAS 21025 + 5221 is a promising candidate of YSOs forming in subcore 1.
There are $\sim20$ WISE sources in the vicinity (Figure \ref{fig:IImaps_IRAS21025}),
suggesting that the subcore has just started to form a small star cluster.
A careful inspection of the C$^{18}$O data and the WISE 12 $\micron$ image indicates that 
the subcore is a part of a small filament extending to the north-east direction.
We delineate its outline by the white broken line in Figure \ref{fig:pv_iras1}(a)
(labeled 1' in the circle).
In addition, the small filament is crossed by another filament
which is an extension of filament 3 in the north-west of the IRAS source.
As can be seen in the channel maps and the PV diagrams shown in panels (b)--(e)
of Figure  \ref{fig:pv_iras1}, these filaments (1' and 3) have different velocities that
can be divided at $V_{\rm LSR}\simeq-1.8$ km s$^{-1}$, and subcore 1 is located 
at their intersection where the 2 velocity components are merged.
These pictures infer that the small filament 1' was collided
by filament 3 in the past, and they were merged at the intersection to form subcore 1.
We suggest that the collision induced the formation of the subcore and then
the small star cluster.

In the case of the other IRAS source (21005+5217), filaments showing such apparent
collisions are difficult to identify. 
Several YSOs including the IRAS source are located between filaments 7 and 12
as shown in Figure \ref{fig:pv_iras2}.
Filament 7 has a branch labeled 7' in the figure
which has the same radial velocity as the main body of the filament
and is extending to the west close to the boundary of filament 12.
As seen in panels (e) of the figure, the branch and filament 12
have slightly different radial velocities by $\sim0.5$ km s$^{-1}$, and their interface exhibits
rather complex velocity field, which might reflect the interaction between the branch of filament 7 and
filament 12. However, it is difficult to confirm the possible interaction and its influence 
on the formation of YSOs only with the present dataset.

To summarize, subcore 1 is very likely to be collided by a part of filament 3, which
may have influenced on the formation of IRAS 21025 + 5221 and the other $\sim20$ YSOs. 
In the case of IRAS 21005+5217,
there are 2 filaments (7 and 12) showing complex velocity field in the vicinity, though it is difficult to 
confirm their possible collision and its influence on star formation.

A clear evidence of collisions between molecular clouds was 
first found in Sgr B \citep{Hasegawa1994, Sato2000}, and since then on,
similar collisions as well as star formation induced by the collisions 
have been evidenced in some star forming regions  \citep[e.g., ][]{Torii2011, Nakamura2014}.
However, these are the collisions between clouds or clumps on much larger scales.
Note that what we suggest in this paper is the collisions between filaments inside 
of a single massive dense core, which may be a direct trigger of formation
of massive stars or star clusters. We will discuss the origin of the colliding filaments
in Section \ref{sec:discussion}.

\section{DISCUSSION \label{sec:discussion}}

To summarize the results of our observations,
L1004E has a huge mass of $\sim 1.1 \times 10^{4}$ $M_\odot$
and it consists of a number of filaments having a mass of $10^2-10^3$ $M_\odot$.
Some of the filaments are apparently colliding against each other,
and it is likely that some YSOs are forming around the regions 
where the filaments are colliding (e.g.,  around IRAS 22025+5221 in Figure \ref{fig:pv_iras1}).

It is probably natural to assume that such collisions induce 
star formation in the filaments. 
Actually, according to recent numerical simulations \citep{Kitsionas2007},
collisions of spherical clumps can yield high SFEs of $10-30$ \%
which is a typical value of  IR clusters \citep{Lada2003}, but head-on collisions
are needed for the high SFEs. Note that collisions of filaments can be partially 
regarded as equivalent to the head-on collisions
of spherical clumps, and formation and collisions of giant filaments
in a massive core like L1004E may play a key role for cluster formation.

Here is a problem, however, to understand the collisions of the filaments.
There have been a number of numerical simulations on the evolution
of dense cores \citep[e.g., ][]{Nakamura2005, Machida2006, Matsumoto2011}. 
In general, the simulations show that filaments can form easily 
in the cores, but they scarcely collide against each other.
The core studied here might have extraordinary parameters (e.g., too high mass or density),
or there might be an unknown mechanism to cause their collisions,
which have not been taken into account in the earlier simulations.

In order to find what physical conditions are needed for 
the collisions of filaments, we performed 
self-gravitational hydrodynamics simulations using the Adaptive Mesh 
Refinement (AMR) technique developed by 
\citet[][]{Matsumoto2007} . 
For the initial conditions, we assumed a uniform core with a huge mass ($1\times 10^4$ $M_\odot$)
and size ($5\times5\times5$ pc$^3$) corresponding to an average density of 
$n_0({\rm H_2})\simeq 1.3\times 10^3$ cm$^{-3}$, which are
similar to those found in L1004E. 
Turbulence typically of an order of Mach 10 ($\sim1.9$ km s$^{-1}$) is imposed
in the initial core \citep[see for detial][]{Matsumoto2011}.  
The initial velocity field is incompressible with a power spectrum of 
$P(k) \propto k^{-4}$, generated following \citet{Dubinski95}, 
where $k$ is the wavenumber.  This power spectrum
results in a velocity dispersion of $\sigma(\lambda) \propto
\lambda^{1/2}$, in agreement with the Larson scaling relations
\citep{Larson81}.  We calculated decaying turbulence with the
self-gravity, and we did not consider a driving force of turbulence
during the evolution \citep[c.f.,][]{Federrath10}.  The gas was
assumed to be isothermal with a temperature of 10 K.  
Evolution of the core was followed from $t =$ 0 Myr to $\sim1$ Myr,
and we observed
the formation and evolution of the resulting filaments
in 2-dimensional velocity channel maps and 3-dimensional 
iso-density maps which are equivalent to those in 
Figures \ref{fig:channelmap_c18o_1} and \ref{fig:filaments}.

Details of the simulations will be presented in a subsequent 
publication (Matsumoto et al. 2014, in preparation).
In this paper, we describe a summary as in the following points (1)--(3),
and show some snapshots of the evolution of the column densities
observed from two different directions ($X$ and $Y$) at some different epochs
in Figure \ref{fig:amr_simulation}(a)--(f): \\
{\bf (1)} Filaments are formed in the core at $t\simeq 0.2$ Myr,
and stars are formed in the filaments spontaneously at $t\simeq 0.6$ Myr.
In total, 209 stars are formed at the end of the simulations ($t\simeq 1$ Myr). \\
{\bf (2)} Apparent collisions of the filaments are not observed
throughout the calculation time.  \\
{\bf (3)} In the velocity channel maps, we sometimes observe a hole and bump 
showing anti-correlations at different velocities, which appears similar to 
those seen in Figures \ref{fig:channelmap_c18o_1} and \ref{fig:collision}(a)--(b).
The anti-correlations observed in the simulations, however, represent relatively
large-scale transient structures drifting in the shear, but they are not 
well-defined filaments forming stars.

We repeated the simulations for several sets of different initial parameters, 
but the results are essentially the same as described in the above.
The holes and bumps on a rather large scale described in the point (3)
may account for a part of the observations, but we could not reproduce 
the well-defined colliding filaments forming stars
by the above simulations, and we concluded that simple increment
of mass and size of the initial core does not result in collisions
of the filaments. An additional condition is apparently 
needed to account for the collisions. 

After some trials, we finally found that the filaments can collide when
there is a large velocity gradient across the initial core in a sense to compress it. 
In other words, we have to force the filaments to collide by giving the velocity gradient.
An example of the simulations incorporating the velocity gradient
can be summarized as in the following point (4):  \\
{\bf (4)} 
If we assume a velocity gradient of $dV/dX \simeq 2$ km s$^{-1}$ pc$^{-1}$ in one direction
across the initial core, which is provided by a sinusoidal flow with an amplitude of Mach 10
\footnote{In addition to the turbulence, we imposed a core-scale
sinusoidal flow of $V(X) = {\cal M} c_s \sin( {2\pi}{\frac{X}{5~\mathrm{pc}}})$ in the $X$-direction on the initial core
for the range $-\frac{5}{2}\leq X \leq \frac{5}{2}$ pc
where ${\cal M} (=10)$ and $c_s(=0.19\,\mathrm {km\,s}^{-1})$ are the Mach number of the flow and
the sound velocity, respectively. This flow provides the maximum velocity
gradient $dV/dX\simeq 2.4$ km s$^{-1}$ pc$^{-1}$ at $X=0$ pc.},
the filaments are formed 
at $t\simeq0.2$ Myr in the same way as described in the point (1),
and some of them start to collide against each other (or cross each other) to form
stars at $t\simeq 0.4$ Myr. 
Stars are often formed at the intersections
of the filaments, but they are also formed spontaneously along the ridge of dense 
filaments without collisions.
Roughly, $25\pm5$ $\%$ of the stars are formed directly by collisions.
In total, 479 stars are formed at the end of the simulation,
which is $\sim2.5$ times larger than in the case of no initial velocity gradient.

Some snapshots of the above simulations are displayed in
Figure \ref{fig:amr_simulation}(g)--(l).
Characteristic features observed in the C$^{18}$O
data, i.e., the anti-correlations in the channel maps
and the velocity jumps in the PV diagrams (Figures \ref{fig:collision}--\ref{fig:pv_iras1}), 
are often seen in the simulated data. In Figure \ref{fig:simulation_color},
we demonstrate an example of such features seen in the simulations. 
Panel (a) of the figure displays the column density distributions 
in different velocity ranges (colored in red and blue) exhibiting
anti-correlations at some places. As shown in panel (b), the first star observed 
in the simulation (indicated by the star symbol) was formed in one of the blue filaments
$\sim 0.1$ Myr after another red filament (indicated by the ellipse with broken line)
collided and passed through it. We can see a clear velocity jump in the PV diagram 
in panel (c) measured along the filaments.
Panels (d) and (e) display the column density distributions in the two different
velocity ranges in panel (a) separately, but the data are smoothed to a lower 
resolution ($0.1$ pc $\simeq 25\arcsec$ at 800 pc) similar to that of the
C$^{18}$O data. Some noticeable anti-correlations including the one around the 
first star mentioned in the above are indicated by boxes, and they appear similar to those
seen in the C$^{18}$O data in Figure \ref{fig:collision}.

It is noteworthy that some elongated condensations in panels (d) and (e) of Figure \ref{fig:simulation_color}
which would be regarded as one filament actually consist of a bunch of thinner filaments that can
be seen at a higher spatial resolution ($\sim0.02$ pc) in panel (a). Filaments observed in the C$^{18}$O 
data may probably have similar substructures which would be resolved by interferometer 
observations with higher angular resolutions of a few arcsec.
Actually, such thin filaments were recently evidenced in the Taurus region
by \citet{Hacar2013} who found bundles of small and gravitationally stable filaments with
a typical length of $\sim0.5$ pc, which may correspond to the thin filaments observed in our
simulations.

As we can see in Figure \ref{fig:amr_simulation}(k),
the point of the simulations is that the velocity gradient generates a layer of high density gas
containing a number of filaments in the middle of the initial core.
In such a layer, star formation should occur more frequently,
not only because the free-fall time should be shorter due to the gas compression,
but also because the collision rate of the filaments should be higher.
Though it is not easy to separate clearly the contributions of the two effects to star formation,
roughly $25\pm5$ $\%$ of the stars are formed directly by collisions of the filaments.
This is consistent with the fact that there are 83 candidates of YSOs selected 
in L1004E (see Section \ref{subset:protostars}) and
$\sim20$ of them are located where collisions of the filaments are inferred
(Figures \ref{fig:pv_iras1} and \ref{fig:pv_iras2}).

In the above simulations, the layer has a thickness of $\sim0.5-1$ pc and 
and its edge-on view appears similar to the
C$^{18}$O intensity map in Figure \ref{fig:ii_maps}(a).
We suggest that this is the case for L1004E.
To be more precise, L1004E may not be a complete edge-on view of the layer
but it may be an oblique view.
In fact, the $^{13}$CO channel maps (Figure \ref{fig:channelmap_13co_1}) shows 
that there is a velocity gradient of $dV/dX\simeq 2$ km s$^{-1}$pc$^{-1}$ 
in the direction orthogonal to the elongation of L1004E
(i.e., from the south-east to north-west, see panels for $-4 < V_{\rm LSR}  < 0$ km s$^{-1}$ 
in the figure). 

For now, it is not yet very clear how large initial velocity gradient is needed 
to cause the collisions of the resultant filaments, but it should be
greater than that necessary to make the crossing time $\tau_{\rm cross}$
shorter than the free fall time $\tau_{ff}=1.05 \left( {\frac{n}{{{{10}^3}~{\rm cm}{^{ - 3}}}}} \right)^{-1/2}$ Myr
of the initial core. In the case of L1004E,
$\tau_{ff}\sim1$ Myr  for the average density $n({\rm H_2})\simeq1.3 \times 10^3$ cm$^{-3}$, and 
$\tau_{\rm cross}\sim 0.5$ Myr if we take $\tau_{\rm cross}$ is the reciprocal of the 
observed velocity gradient $dV/dX$ ($\simeq 2$ km s$^{-1}$pc$^{-1}$).

It is not clear either how the initial velocity gradients influence on the final SFEs 
of the cores, because we have results only for the two cases 
with ($dV/dX\simeq2$ km s$^{-1}$pc$^{-1}$) and without ($dV/dX=0$ km s$^{-1}$pc$^{-1}$) 
the velocity gradients at the moment. 
We would expect higher SFEs for larger velocity gradients because the initial cores can be 
compressed more sufficiently,
but it should be clarified by additional simulations.

The velocity gradient necessary for collisions of the resultant filaments
can be caused by an external effect, e.g., shock fronts of SNRs 
or H{\scriptsize{II}} regions to compress the initial core. Actually, there have been 
a number of studies for such molecular clouds influenced by
SNRs \citep[e.g.,][]{Tatematsu1987,Tatematsu1990, Moriguchi2001}
or H{\scriptsize{II}} regions  \citep[e.g.,][]{Dobashi2001, Toujima2011, Shimoikura2013, Chibueze2013}.
It is noteworthy that the Cyg OB7 molecular cloud has been suggested to be 
interacting with the nearby SNR HB21 \citep[cataloged as G89.0+4.7 by][]{Green1998} by \citet{Tatematsu1990}.
In addition, there are some other SNRs (i.e., DA530, DA551, and 3C434.1) 
in the vicinity of the Cyg OB7 cloud 
\citep[][see their Figure 11]{Dobashi1994}. Although distances to these SNRs
have not been determined well \citep[e.g., ][]{Mavromatakis2007},
it is very possible that they have influenced on the initial velocity fields of L1004E.
Other than the influence of the SNRs and H{\scriptsize{II}} regions, stellar winds from nearby OB stars can also
provide a similar effect on the velocity field of the initial cores.  Note that there have been found several
young OB stars located $5-10$ pc away from L1004E \citep{Reipurth2008}.

In Figure \ref{fig:illustration}, we summarize the 
suggested scenario for the formation of the colliding filaments. 
The collision of the filaments in massive cloud cores may be essential
for cluster formation. The ubiquitousness of the scenario, however,
has to be confirmed especially by observing other massive star forming regions or
cluster forming regions such as M17 and Ori KL, because, at the moment, L1004E is 
probably only the core where such collisions of giant filaments are found.


\section{CONCLUSIONS \label{sec:conclusions}}

We have carried out millimeter-wave observations of a massive dense core L1004E 
in the Cyg OB 7 molecular cloud 
in various molecular lines such as $^{12}$CO,  $^{13}$CO, C$^{18}$O, CCS, 
HC$_3$N, and NH$_3$, using the 45m telescope at the Nobeyama Radio Observatory.
Main findings of this paper are summarized below. 

(1) The molecular observations revealed the total extent and velocity structures of L1004E.
Based on the C$^{18}$O data, we find that the core has a huge mass 
and size of $1.1 \times 10^4$ $M_\odot$ and $\sim 5\times2$ pc$^2$, respectively. 
The core is also characterized by the cold temperature $\sim10$ K as measured in the
$^{12}$CO and NH$_3$ molecular lines.
The maximum column density observed is $N({\rm H_2})\simeq5\times 10^{22}$ cm$^{-2}$
at the peak position of the C$^{18}$O intensity map.
Turbulent motion measured from the line width varies in the range 
$\Delta V {\rm( C^{18}O)} \simeq 1-3$ km s$^{-1}$ over the core.

(2) We searched for candidates of YSOs in the IRAS PSC, and found
that there are only 2 sources that can be regarded as promising YSOs. 
They are IRAS 21005+5217 and 21025+5221. We employed a stellar model
developed by \citet{Robitaille2007} to access the age, mass, and luminosity of
the IRAS sources.  
We searched for YSOs also in the WISE PSC which is much more sensitive
than the IRAS PSC by using 
the selection criteria proposed by \citet{Koenig2012}.
If we disregard one of their criteria to exclude AGNs,
there found 83 candidates for YSOs in the observed region,
and $\sim20$ of them are concentrated in a condensation
around IRAS 21025+5221.
Assuming that all of the WISE sources are real YSOs,
we estimated the star formation efficiency of the entire core
to be ${\rm SFE}\lesssim 1$ \% at most.

(3) We find that the core consists of a number of filaments and some
core-like structures.
In the velocity channel maps of C$^{18}$O,
these filaments appear apparently colliding against each other,
and some candidates of YSOs are located near the intersections of the filaments.
We identified 12 major filaments and core-like condensations, and estimated their masses to be 
$10^2-10^3$ $M_\odot$. We investigated their gravitational stability
to find at least half of the filaments and core-like condensations are likely to be in the Virial equilibrium,
and the others might be gravitationally unbound.

(4) Using the adaptive mesh regiment technique \citep{Matsumoto2007},
we performed numerical simulations to reproduce the collisions of the 
filaments. As the initial conditions of the core, we assumed a high mass, 
size, and density similar to those observed toward L1004E.
Results of the simulations indicate that the filaments are formed in the core, 
but they never collide against each other during their evolution. 
After some trials, we finally found that the filaments can collide only
when there is a large velocity gradient in the initial core in a sense to
compress it, which can be generated by an external energetic effect 
such as the shock fronts of SNRs. There are actually some SNRs in the 
vicinity of the Cyg OB 7 molecular cloud including HB21
that has been suggested to be interacting with the cloud.
We suggest that L1004E was influenced by 
such an external effect (possibly by HB21) to have the initial 
velocity gradient, which results in the formation of the colliding 
filaments.


\acknowledgments
This research was financially supported by Grant-in-Aid for Scientific Research
(Nos. 22340040, 23540270, 23540270, 24244017, 24700866, 26287030, 26350186, 26400233, and 26610045)
of Japan Society for the Promotion of Science (JSPS), and also by the Mitsubishi Foundation. The 45m radio telescope is operated by NRO, a branch of National Astronomical Observatory of Japan. 
This research has made use of the NASA/ IPAC Infrared Science Archive, which is operated by the Jet Propulsion Laboratory, California Institute of Technology, under contract with the National Aeronautics and Space Administration.
The Wide-Field Infrared Survey Explorer is a joint project of the
University of California, Los Angeles, and the Jet Propulsion
Laboratory (JPL), California Institute of Technology (Caltech),
funded by the National Aeronautics and Space Administration
(NASA).
The Two Micron All Sky Survey is a joint project of the
University of Massachusetts and the Infrared Processing and
Analysis Center/California Institute of Technology, funded by
the National Aeronautics and Space Administration and the
National Science Foundation.

\clearpage


\appendix

\section{Derivation of the Molecular Column Density and the Mass of the Core}

We derived the total mass of the observed core as summarized in the following.
The method we used assumes the LTE, and is a standard way to derive
the total molecular mass from the C$^{18}$O emission line.
Details of the method can be found in some literature  \citep[e.g., ][]{Shimoikura2012}.

\begin{figure*}
\begin{center}
\includegraphics[scale=0.8]{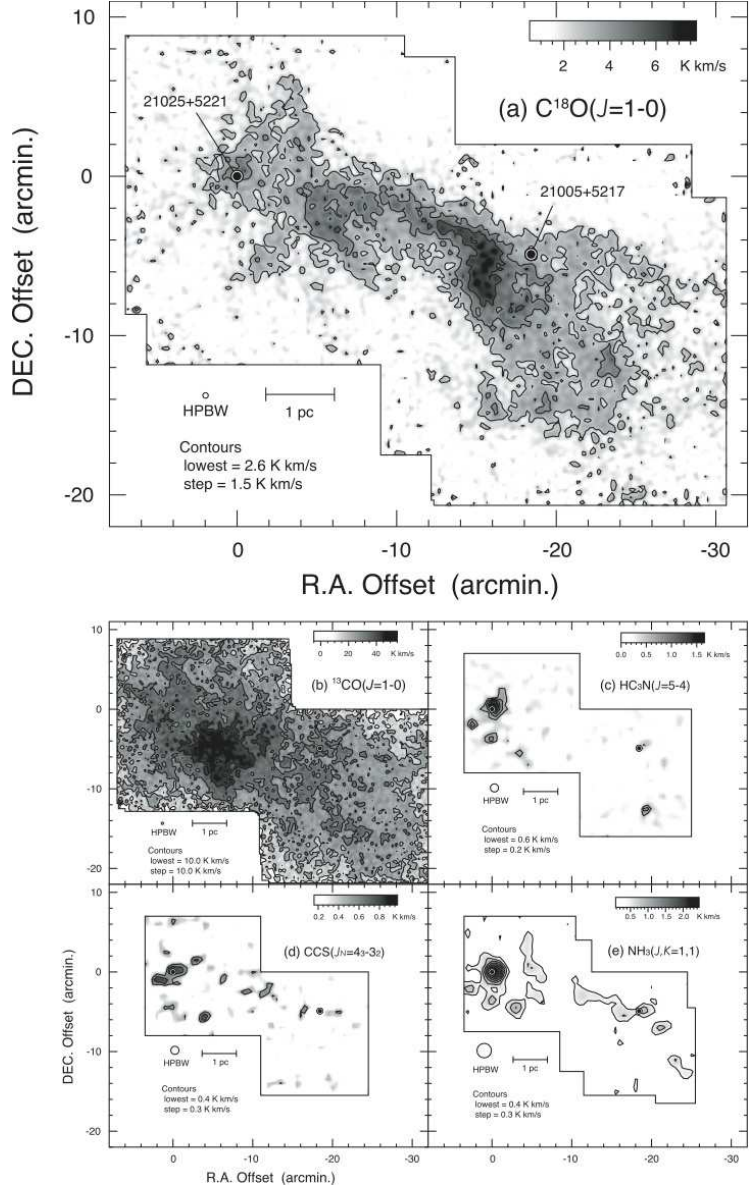}
\caption{
Integrated intensity distributions of the 
(a) C$^{18}$O$(J=1-0)$,
(b) $^{13}$CO$(J=1-0)$,
(c) HC$_3$N$(J=5-4)$, 
(d) CCS$(J_N=4_3-3_2)$, and 
(e) NH$_3(J,K=1,1)$ emission lines
observed with the 45m telescope. 
Thin solid lines denote the observed areas.
The $^{13}$CO map is integrated over the velocity range $-15<V_{\rm LSR}<6$ km s$^{-1}$,
and the other maps are integrated over the range  $-4<V_{\rm LSR}<0$ km s$^{-1}$.
The maps are shown as an offset from IRAS 21025+5221 (Table \ref{tab:iras}).
\label{fig:ii_maps}}
\end{center}
\end{figure*}

\begin{figure}
\begin{center}
\includegraphics[scale=1.0]{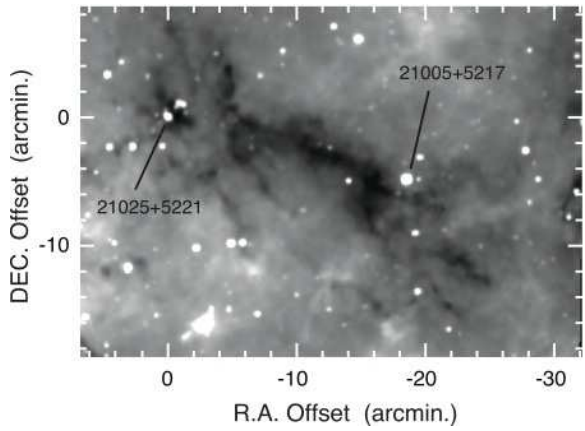}
\caption{
The WISE 12 $\micron$ image of L1004E.
Positions of the two IRAS sources are indicated (see Section \ref{subset:protostars}).
The map is shown as an offset from IRAS 21025+5221.
\label{fig:w3}}
\end{center}
\end{figure}

\begin{figure}
\begin{center}
\includegraphics[scale=1.0]{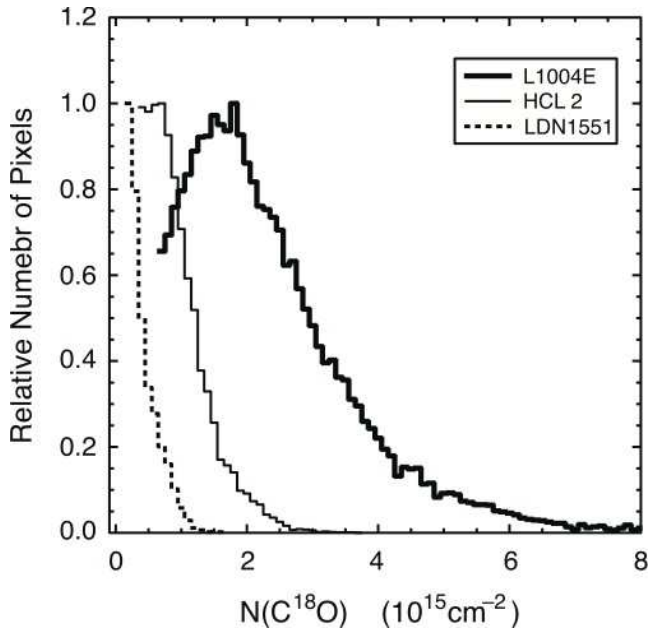}
\caption{
Frequency distributions of the C$^{18}$O column densities of 
L1004E (thick solid line), HCL 2 (thin solid line), and LDN 1551 (broken line)
observed with the 45m telescope. The maximum number of pixels in the vertical 
axis is normalized to unity for each region.
\label{fig:column_density}}
\end{center}
\end{figure}

\begin{figure*}
\begin{center}
\includegraphics[scale=0.8]{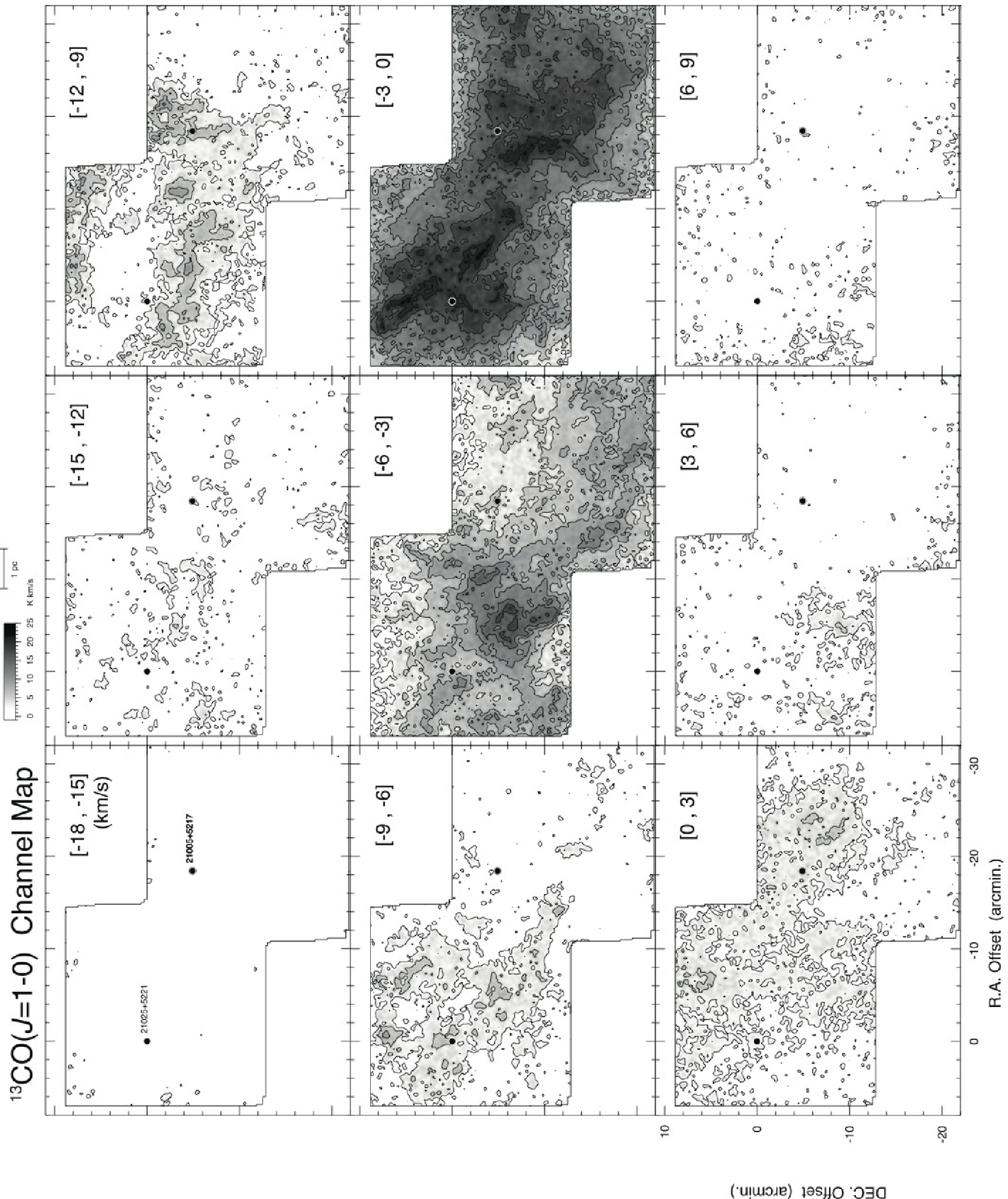}
\caption{
Channel maps of  the $^{13}$CO$(J=1-0)$ emission line.
Numbers in the parentheses at the top-right corner of each panel
indicate the velocity ranges used for the integration.
Contours start from 2 K km s$^{-1}$ with a step of 3 K km s$^{-1}$.
\label{fig:channelmap_13co_0}}
\end{center}
\end{figure*}

\begin{figure*}
\begin{center}
\includegraphics[scale=0.8]{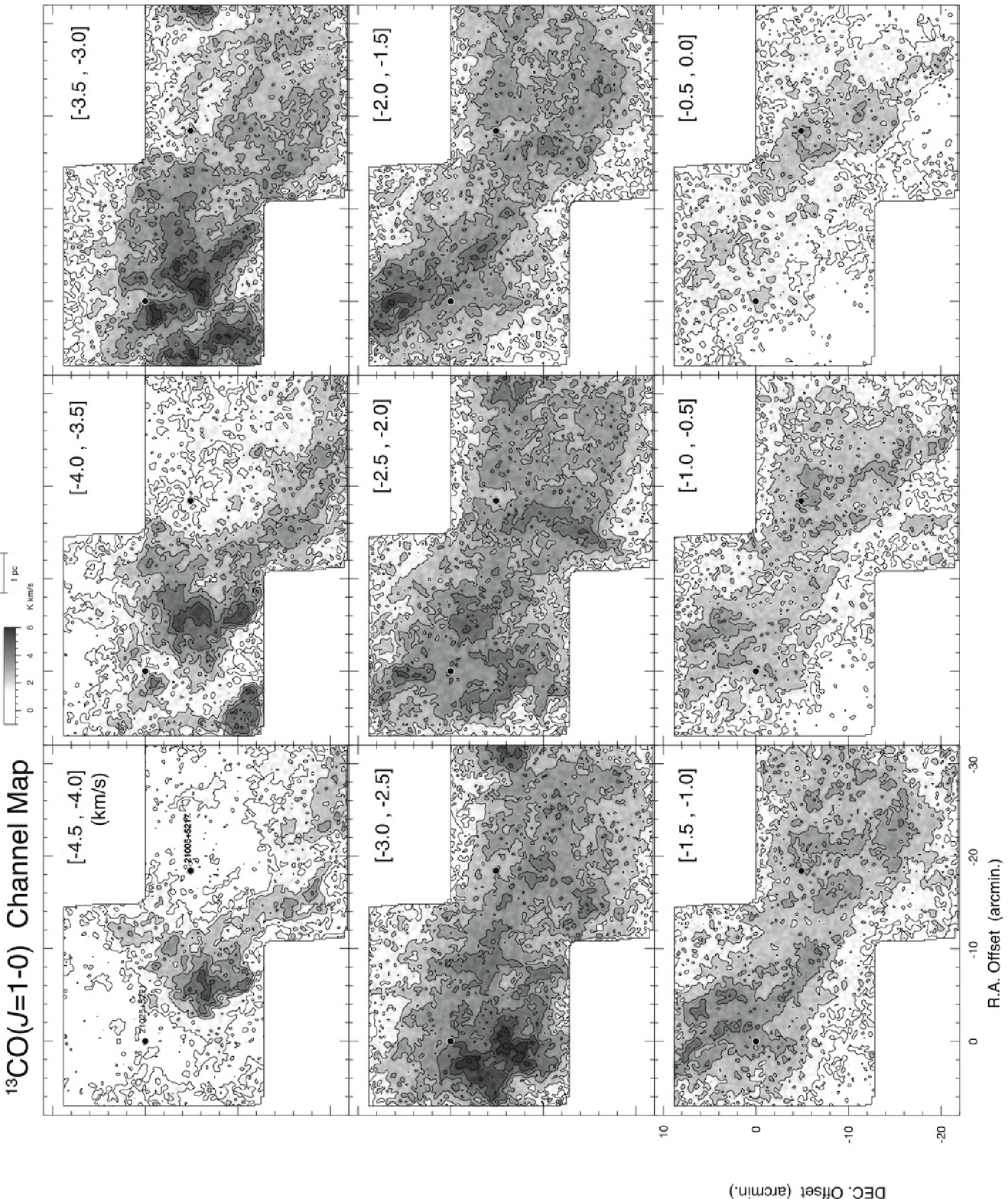}
\caption{
Channel maps of  the $^{13}$CO$(J=1-0)$ emission line
with a higher velocity resolution (0.5 km s$^{-1}$).
Numbers in the parentheses at the top-right corner of each panel
indicate the velocity ranges used for the integration.
Contours start from 1 K km s$^{-1}$ with a step of 1 K km s$^{-1}$.
\label{fig:channelmap_13co_1}}
\end{center}
\end{figure*}

\begin{figure*}
\begin{center}
\includegraphics[scale=0.8]{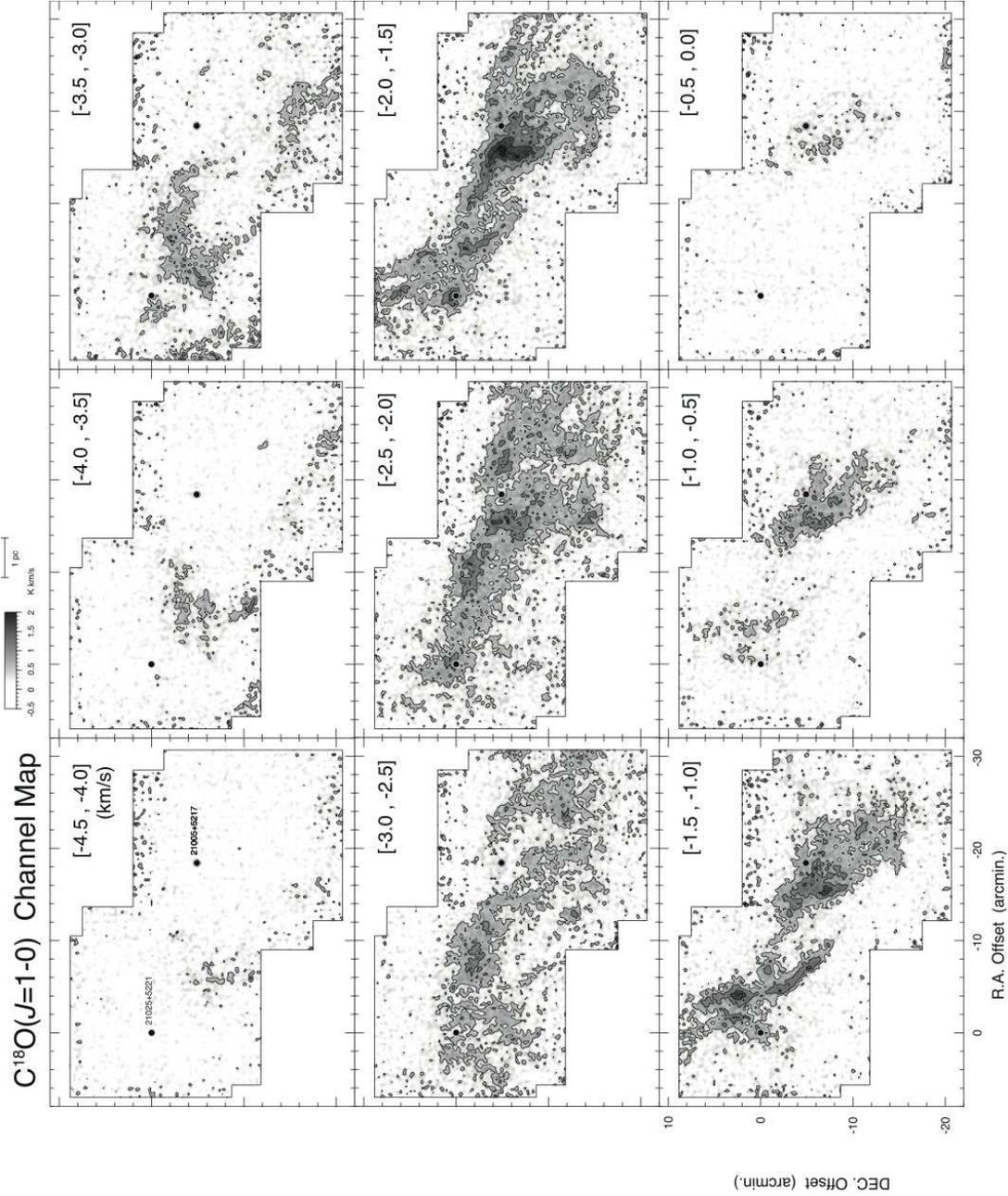}
\caption{
Channel maps of  the C$^{18}$O$(J=1-0)$ emission line
for the same velocity ranges as in Figure \ref{fig:channelmap_13co_1}.
In each panel, the lowest contours and contour intervals are 0.5 K km s$^{-1}$.
\label{fig:channelmap_c18o_1}}
\end{center}
\end{figure*}

\begin{figure}
\begin{center}
\includegraphics[scale=1.0]{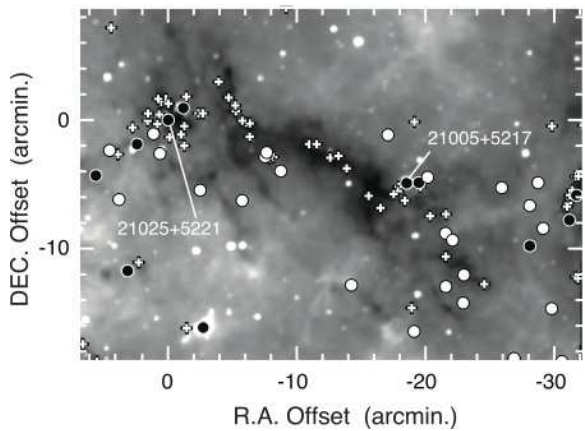}
\caption{
Distributions of the WISE point sources shown on the 
WISE 12 $\micron$ image.
Class I and Class II sources selected following
the criteria of \citet{Koenig2012} are shown
by the filled and open circles, respectively.
Small plus signs indicate the sources which would be 
regarded as YSOs (Class I or II) without one of their criteria
to exclude ANGs
(see Section \ref{subset:protostars}). Point sources
which are likely to be false detections in the original WISE PSC without apparent counterparts
in none of the four band WISE images (i.e., 3.4, 4.6, 12, and 22 $\micron$) are excluded.
\label{fig:ysos}}
\end{center}
\end{figure}

\begin{figure}
\begin{center}
\includegraphics[scale=1.0]{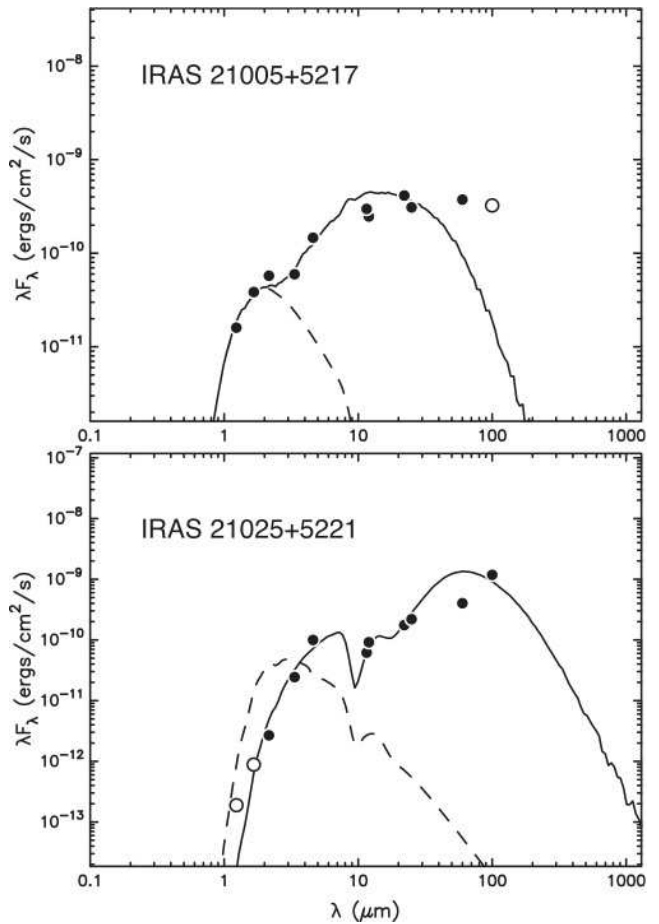}
\caption{Spectral energy distributions of IRAS 21005+5217 (the top panel) and 21025+5221 (the bottom panel).
The filled circles show the observed fluxes summarized in Tables \ref{tab:iras} and \ref{tab:counterparts}. 
The open circles denote the upper limits.
The solid line is the model of \citet{Robitaille2007} best fitting the observed fluxes,
which was calculated using their tools available on the web (http://caravan.astro.wisc.edu).
The dashed line shows the stellar photosphere of the central star best fitting the observed data, 
as we would observe in the absence of circumstellar dust but including interstellar extinction.
\label{fig:sed}}
\end{center}
\end{figure}

\begin{figure*}
\begin{center}
\includegraphics[scale=0.6]{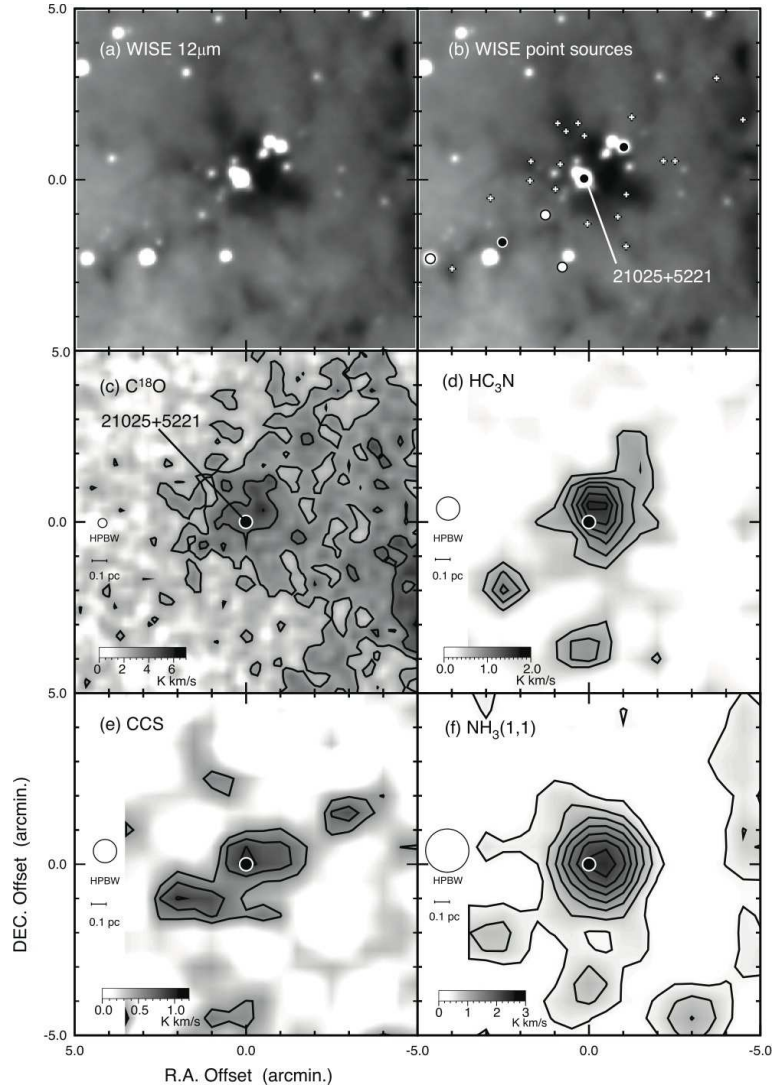}
\caption{
Images of the condensation around IRAS 21025+5221.
Panel (a) shows the WISE 12 $\micron$ image, and panel (b) shows the 
distributions of WISE point sources in the same way as in Figure \ref{fig:ysos}.
Other panels show the integrated intensity
distributions of the
(c) C$^{18}$O$(J=1-0)$,
(d) HC$_3$N$(J=5-4)$, 
(e) CCS$(J_N=4_3-3_2)$, and 
(f) NH$_3(J,K=1,1)$ emission lines.
Contours are the same as in Figure \ref{fig:ii_maps}.
Position of IRAS 21025+5221 is indicated by a larger filled circle in panels (c)--(f).
\label{fig:IImaps_IRAS21025}}
\end{center}
\end{figure*}

\begin{figure*}
\begin{center}
\includegraphics[scale=1.0]{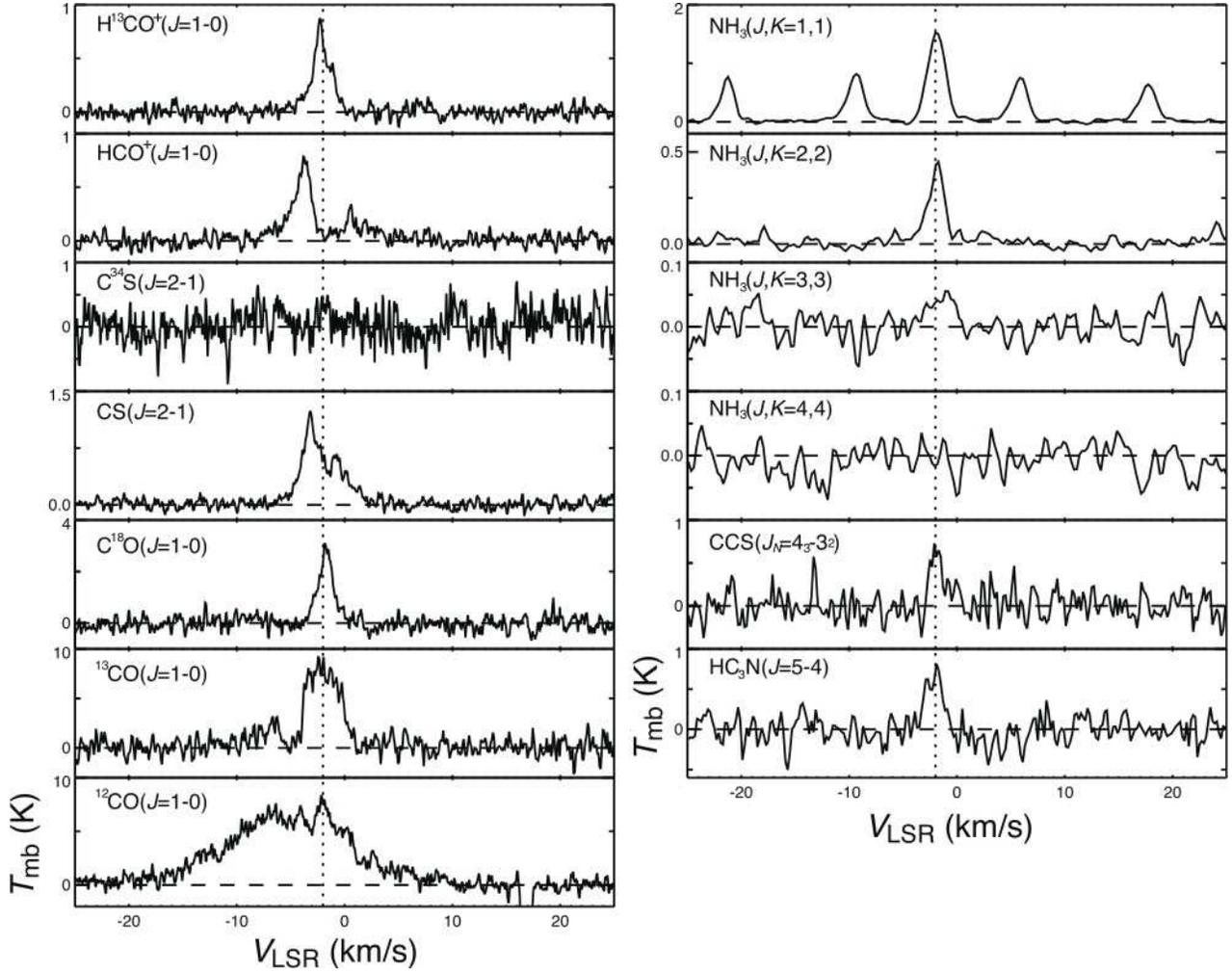}
\caption{
Various molecular lines observed toward IRAS 21025+5221.
The broken line denotes $V_{\rm LSR}=-2$ km s$^{-1}$.
\label{fig:Sample_spectra}}
\end{center}
\end{figure*}

\begin{figure}
\begin{center}
\includegraphics[scale=1.0]{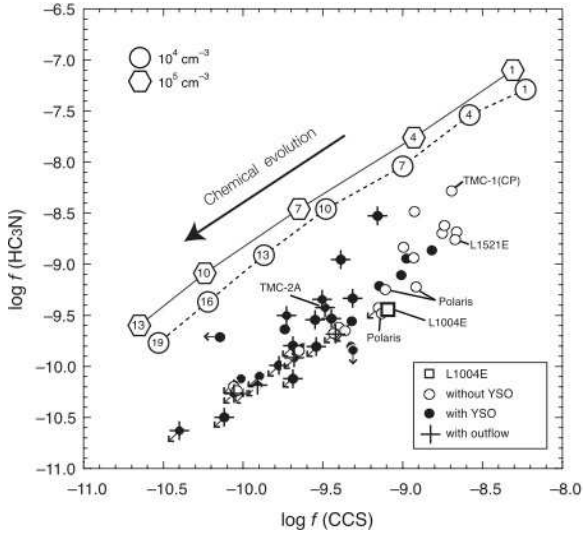}
\caption{
Fractional abundance of HC$_3$N versus that of CCS
observed toward L1004E (square) compared with
those in other dense cores in literature such as TMC-1(CP) and TMC-2A
corrected by \citet{Shimoikura2012} who studied starless cores in Polaris. Open and filled circles denote
cores with and without YSOs, and those accompanied by molecular outflows
are indicated by plus signs. Arrows denote the upper limits.
Results of model calculations performed by \citet{Suzuki1992} for 
densities of $n$(H$_2$)=10$^4$ and $10^5$ cm$^{-3}$ are shown for 
comparison by open circles and hexagons. Numbers in the symbols
denote the reaction time in units of 10$^5$ yr.
\label{fig:abundances}}
\end{center}
\end{figure}

\begin{figure}
\begin{center}
\includegraphics[scale=1.0]{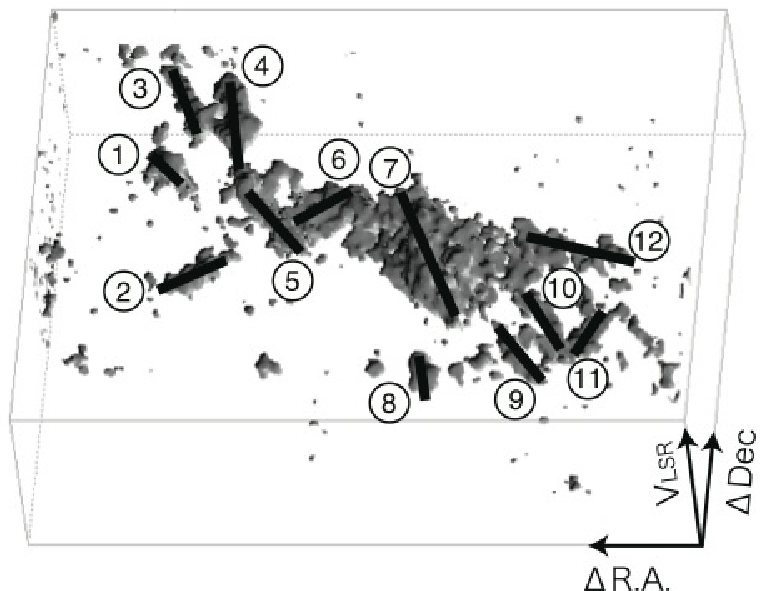}
\caption{
Locations of the 12 filaments and subcores in Table \ref{tab:filaments} shown in the 3 dimensional iso-temperature map 
of $T_{\rm mb}($C$^{18}$O$)=2$ K.
\label{fig:filaments}}
\end{center}
\end{figure}

\begin{figure*}
\begin{center}
\includegraphics[scale=1.0]{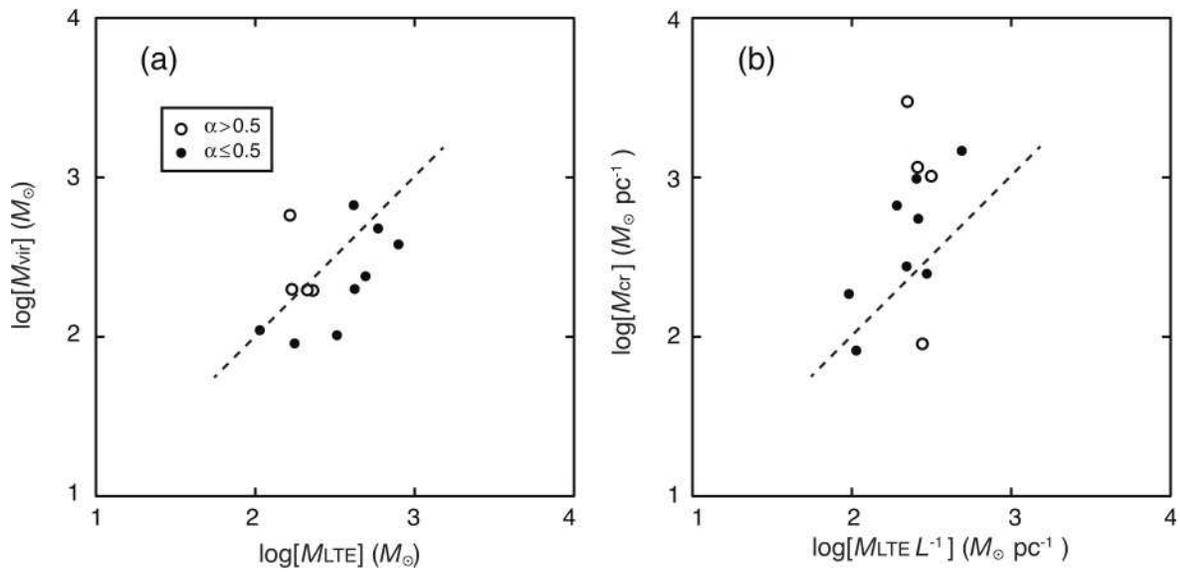}
\caption{
(a) Virial mass vs. LTE mass of the filaments, and 
(b) critical mass vs. LTE mass per unit length of the filaments.
Broken lines denote the equality.
Open circles represent rather round filaments with a higher ellipticity of $\alpha > 0.5$
($\alpha =1$ is for a complete sphere)
which we call ``subcores" in Section \ref{subset:filaments}, and filled circles denote
the other elongated filaments with $\alpha \le 0.5$.
\label{fig:Mvir}}
\end{center}
\end{figure*}

\begin{figure*}
\begin{center}
\includegraphics[scale=1.0]{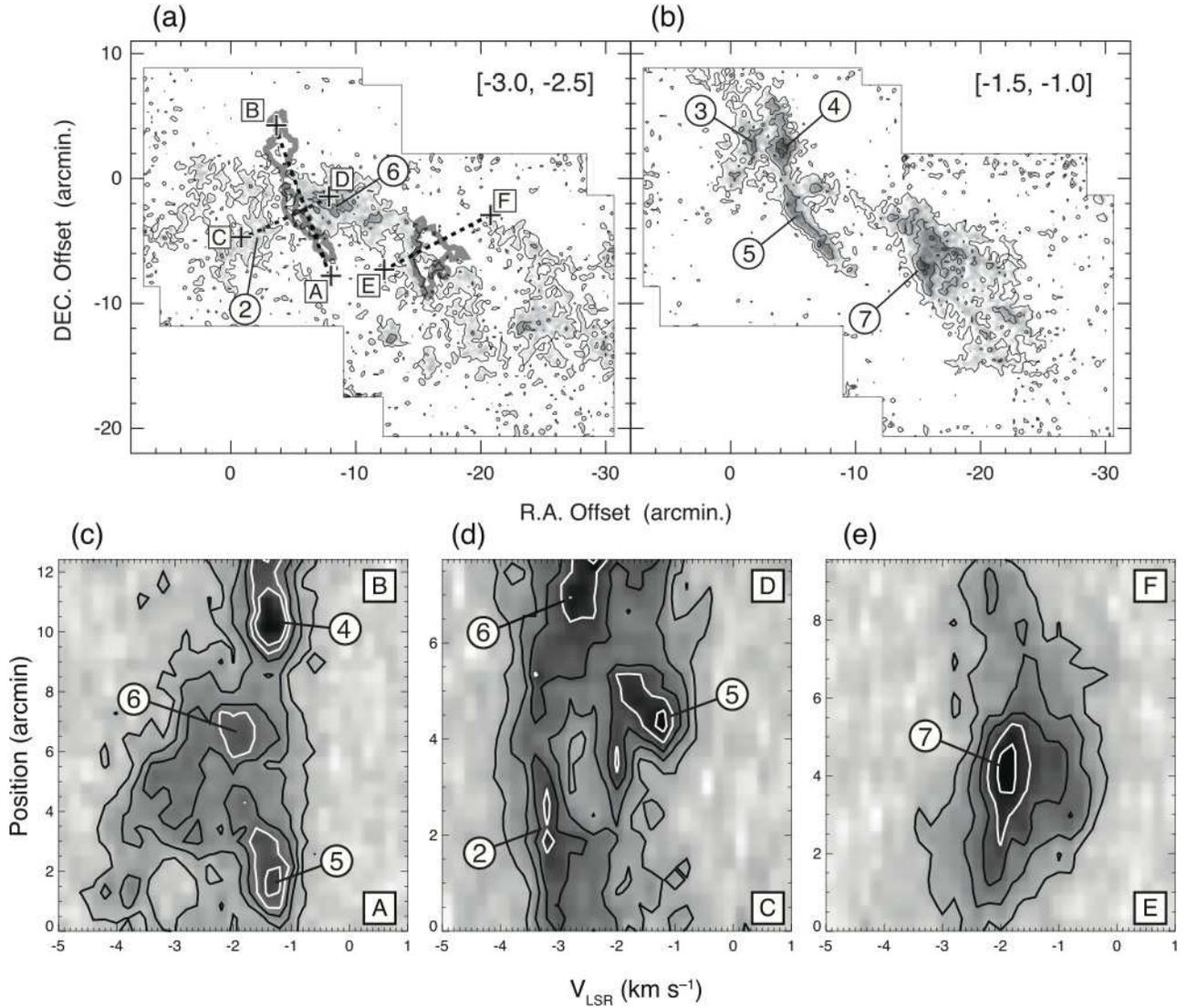}
\caption{
Panels (a) and (b) display two channel maps in Figure \ref{fig:channelmap_c18o_1} showing
apparent anti-correlations. Numbers 2--7 in circles denote the identified
filaments and subcores in Table \ref{tab:filaments}. Some contours of the filaments No. 4, 5, and 7
in panel (b) are indicated by thick gray lines in panel (a) for comparison.
Plus signs labeled  A--F in squares in panel (a) denote the positions where position-velocity (PV)
diagrams displayed in panels (c)--(e) are measured.
The lowest contours and contour intervals of the PV diagrams are $T_{\rm mb}({\rm C^{18}O})=$0.4 K for
the diagrams in panels (c) and (d), and 0.6 K for the diagram in panel (e).
\label{fig:collision}}
\end{center}
\end{figure*}

\begin{figure*}
\begin{center}
\includegraphics[scale=0.8]{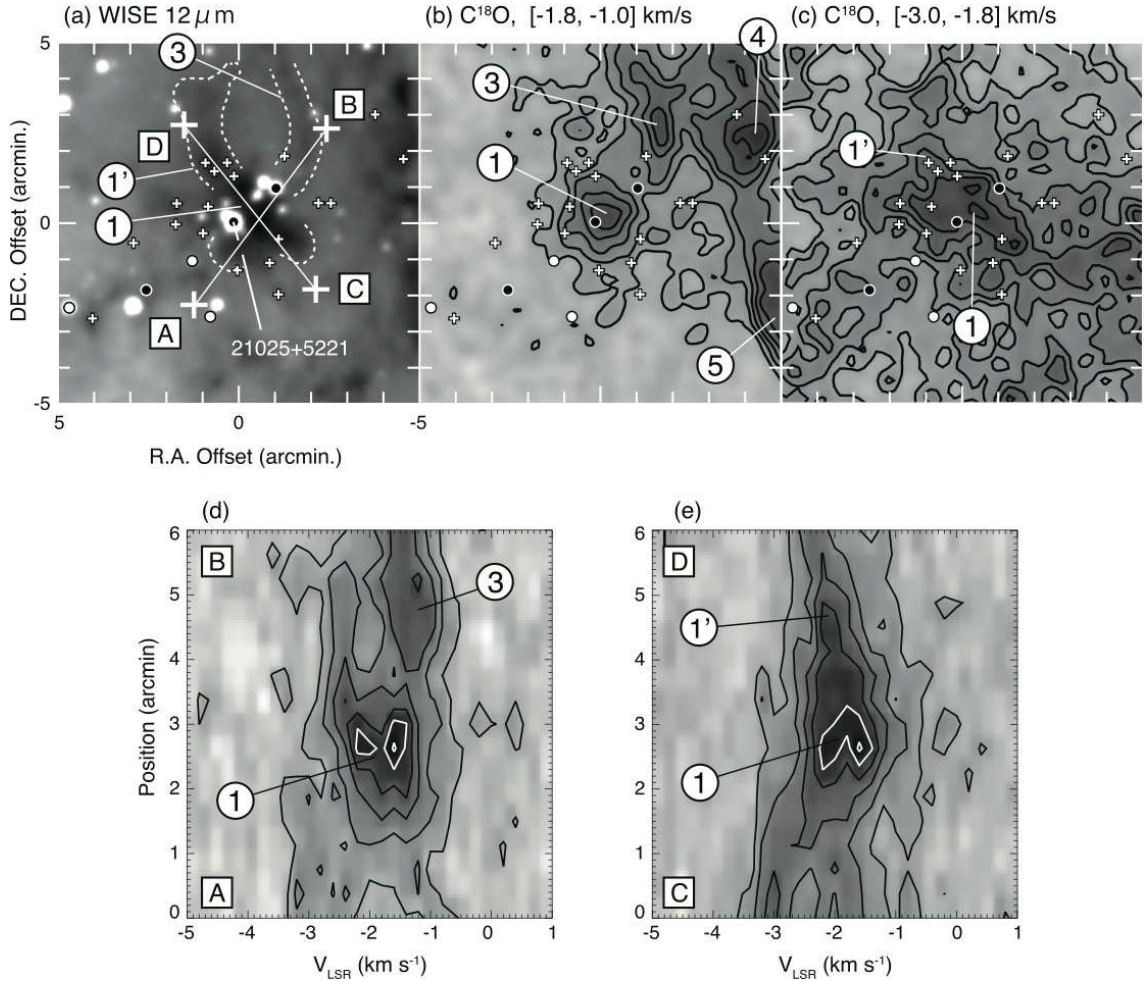}
\caption{
Distributions of YSOs and the C$^{18}$O emission
around IRAS 21025+5221. Panel (a) shows the WISE 12 $\micron$ image, and
panels (b) and (c) show the  C$^{18}$O intensity integrated over the velocity ranges
indicated above each panel. YSOs are shown in the same way as in Figure \ref{fig:ysos}.
Numbers in circles denote the identified
filaments and subcores.
White broken lines in panel (a) delineate the outlines
of some filaments and subcores that can be identified in the WISE image. 
Plus signs labeled  A--D in squares denote the positions where PV diagrams
displayed in panels (d) and (e) are measured.
The lowest contours and contour intervals of the PV diagrams are $T_{\rm mb}({\rm C^{18}O})=$0.4 K.
\label{fig:pv_iras1}}
\end{center}
\end{figure*}

\begin{figure*}
\begin{center}
\includegraphics[scale=0.8]{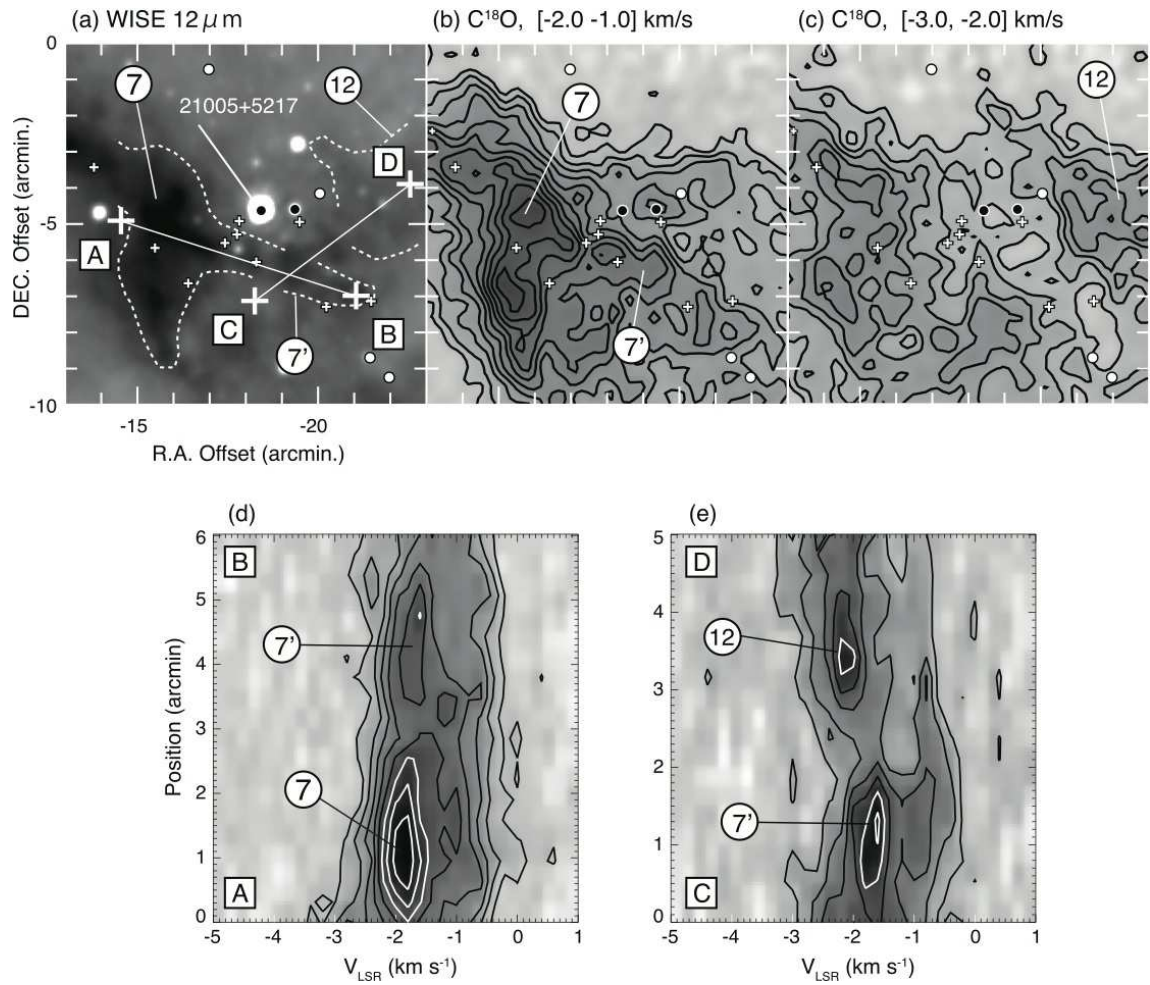}
\caption{Same as Figure \ref{fig:pv_iras1}, but for IRAS 21005+5217.
\label{fig:pv_iras2}}
\end{center}
\end{figure*}

\begin{figure*}
\begin{center}
\includegraphics[scale=0.8]{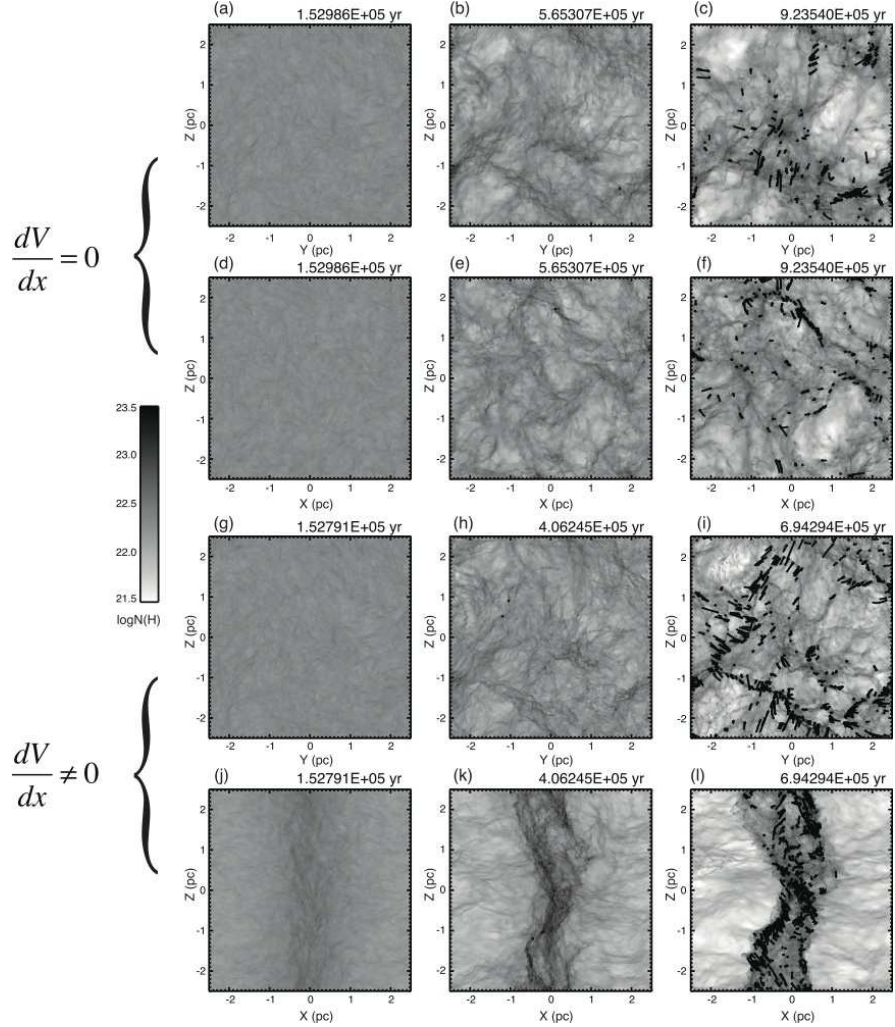}
\caption{
Snapshots of the simulations with and without the initial velocity gradient. 
The upper six panels (a)--(f) show the time evolution of the total column density for the case 
of no velocity radiant. Panels (a)--(c) show the column density observed 
along the $X$ axis, and panels (d)--(f) show the column density observed along the $Y$ axis. 
Numbers above the panels represent passage of time in the simulation.
The gas density is set to be uniform in the beginning ($t=0$ yr).
Solid lines indicate the proper motions of the formed stars.
Panels (a) and (d) are the snapshots at an early stage of the core evolution, 
panels (b) and (e) are those just after the formation of the first stars, 
and panels (c) and (f) are those close to the end of the simulation.
The lower six panels (g)--(l) are for the case with the initial velocity gradient
imposed along the $X$ axis ($dV/dX\simeq2$ km s$^{-1}$ pc$^{-1}$), and they are displayed 
in the same way as panels (a)--(f).
\label{fig:amr_simulation}}
\end{center}
\end{figure*}

\begin{figure*}
\begin{center}
\includegraphics[scale=0.8]{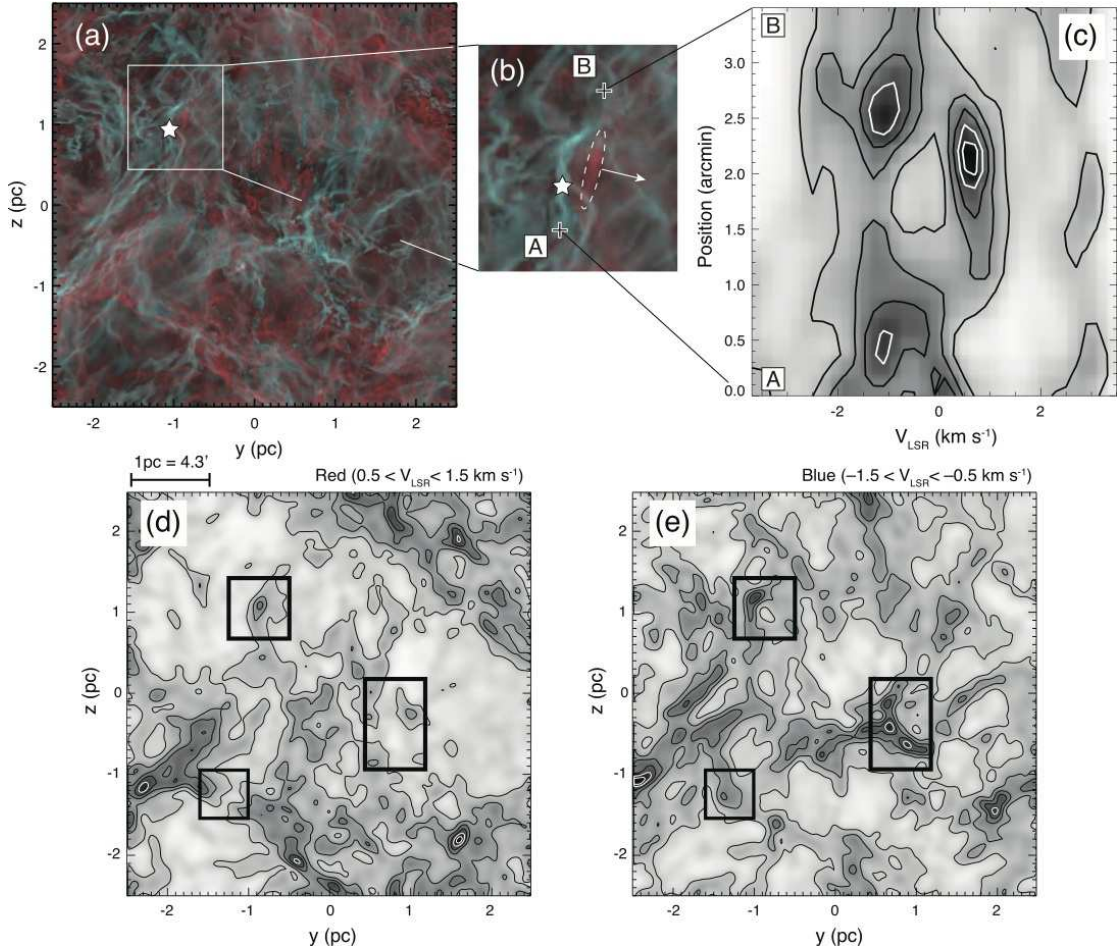}
\caption{
(a) An example of the column density distributions 
observed in the simulation shown in Figure \ref{fig:amr_simulation}(h).
Blue and red colors express the column densities contained
in the velocity ranges $-1.5<V<-0.5$ km s$^{-1}$
and $0.5<V<1.5$ km s$^{-1}$, respectively. Star symbol indicates
the first star observed in the simulation. The original spatial resolution
of the simulation is $\sim0.005$ pc, which is smoothed to $\sim0.02$ pc
in the image.
(b) Close-up view of the image in panel (a).
The first star was formed in the blue filament $\sim0.1$ Myr after the red filament 
(indicated by the ellipse with broken line) passed through the blue filament.
The arrow denotes the movement of the red filament on the YZ plane.
(c) Position-Velocity diagram measured along the line A--B in panel (b).
The lowest contours and contour intervals are 2 $\times 10^{21}$ cm$^{-2}$(km s$^{-1}$)$^{-1}$.
Unit of the vertical axis is converted to arcmin for the distance of L1004E (800 pc).
The data are smoothed to the 0.2 km s$^{-1}$ and 0.1 pc ($\simeq25\arcsec$) resolutions
similar to those of the C$^{18}$O data.
(d) Column density distributions contained in the velocity range $-1.5<V<-0.5$ km s$^{-1}$
same as the blue component in panel (a), but the image is smoothed to the 0.1 pc resolution.
The lowest contours and contour intervals are 0.3 $\times 10^{21}$ cm$^{-2}$.
(e) Same as panel (d), but for the red component in the velocity range
$0.5<V<1.5$ km s$^{-1}$ in panel (a). Boxes in panels (d) and (e) denote the regions
where anti-correlations are seen. 
\label{fig:simulation_color}}
\end{center}
\end{figure*}

\begin{figure*}
\begin{center}
\includegraphics[scale=0.7]{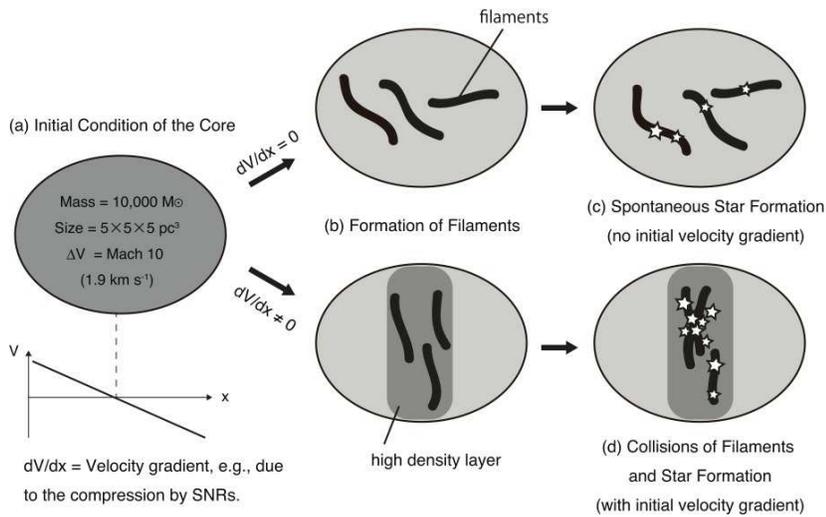}
\caption{
Schematic illustration of evolution of L1004E
inferred from simulations. 
(a) Initial stage of the core with uniform density, which is the starting point of the simulations ($t=0$ Myr). 
The total mass, size, turbulence are taken from the observations. We performed 2 types of simulations: 
In one case, we assumed no velocity gradient in the core, but in the other case, we assumed a  
velocity gradient ($dV/dX\simeq2$ km s$^{-1}$ pc$^{-1}$) in a sense to compress the core in one direction.
(b) The stage when a number of filaments are formed by the self-gravity, but no star has been 
formed yet ($t\simeq$0.2 Myr). A layer with high density is formed when there is the initial velocity gradient,
and the filaments are more or less aligned along the layer (lower panel).
(c) If there is no initial velocity gradient, stars are formed in the filaments spontaneously ($t \simeq 0.6$ Myr).
(d) If there is the initial velocity gradient, some filaments collide against each other before or after the
star formation ($t \simeq 0.4$ Myr), and more stars are formed than in the case without the initial velocity gradient.
\label{fig:illustration}}
\end{center}
\end{figure*}

In general, observed brightness temperature of a certain molecular line
$T_{\rm mb}$($X$) (e.g., $X=$C$^{18}$O, $^{12}$CO, etc.)  can be expressed as 
\begin{equation}
\label{eq:radi_transfer}
{T_{\rm mb}}(X) = \left[ {J({T_{\rm ex}}) - J({T_{\rm bg}})} \right]\left[ {1 - {{\mathop{\rm e}\nolimits} ^{ - \tau (X)}}} \right]
\end{equation} 
where $J(T) = \frac{{{T_0}}}{{\exp ({T_0}/T) - 1}}$ , and $T_0$ is
a constant ( $T_0$=5.27 K for C$^{18}$O and 5.53 K for $^{12}$CO).
$T_{\rm ex}$ and  $T_{\rm bg}$ are the excitation temperature and 
the cosmic background (2.7 K), respectively, and $\tau (X)$ is
the optical depth.
We first estimated $T_{\rm ex}$ at each observed position
by solving the above equation for $T_{\rm mb}$($^{12}$CO) measured at the peak 
velocity of the C$^{18}$O spectra assuming $\tau$($^{12}$CO)$>>1$.
We then derived $\tau$(C$^{18}$O) and $N$(C$^{18}$O), and 
converted $N$(C$^{18}$O) to $N({\rm H_2})$ to estimate
the total mass $M_{\rm core}$ by summing up the derived $N({\rm H_2})$
over the observed area. The process of the derivation 
can be summarized in the following equations;

\begin{equation}
\label{eq:tau}
\tau \left( {{\rm C^{18}O}} \right) =  - \ln \left[ {1 - \frac{{{T_{\rm mb}}({\rm C^{18}O})}}{{J({T_{\rm ex}}) - J({T_{\rm bg}})}}} \right] 
\end{equation} 
\begin{equation}
\label{eq:N18}
N({\rm C^{18}O}) = \frac{{C_0}{T_{\rm ex}}}{{1 - \exp ( - \frac{{T_0}}{{{T_{ex}}}})}}\int {\tau ({\rm C^{18}O})dv}   
\end{equation} 
\begin{equation}
\label{eq:NH2}
N({\rm H_2}) = \left[ {\frac{{N({\rm C^{18}O})}}{{1.8 \times {{10}^{14}}}} + 3.7} \right] \times {10^{21}}
\end{equation} 
\begin{equation}
\label{eq:mass}
{M_{\rm core}} = \mu {m_{\rm H}}S_{\rm pix}\sum {N({\rm H_2})}
\end{equation} 
where $N({\rm C^{18}O})$ and $N({\rm H_2})$ are in units of cm$^{-2}$, and
the constant $C_0$ is $2.52\times 10^{14}$ cm$^{-2}$ K$^{-1}$ (km s$^{-1}$)$^{-1}$ for the C$^{18}$O$(J=1-0)$ emission line.
$\mu$, $m_{\rm H}$, and $S_{\rm pix}$ in the last equation are the mean molecular weight taken to be 2.4,
the hydrogen mass, and the area of the pixels of the C$^{18}$O 
map ($10{\arcsec} \times 10{\arcsec} \simeq 1.432 \times 10^{34}$ cm$^2$ at the assumed distance 800 pc).
Summation in the right side of Equation (\ref{eq:mass}) is made for the total area 
observed in C$^{18}$O shown in Figure {\ref{fig:ii_maps}}(a).
Equation (\ref{eq:NH2}) is derived from an empirical relationship between 
$A_V$ and $N$(C$^{18}$O) \citep{Frerking1982}  as well as between
$A_V$ and $N$(H$_2$) for $R_V=3.1$ \citep{Bohlin1978}.

Note that $N({\rm H_2})$ calculated using Equation (\ref{eq:NH2}) and then
$M_{\rm core}$ calculated using Equation (\ref{eq:mass}) may be the minimum estimates for the true values,
because C$^{18}$O is known to be depleted onto dust grains in very dense cloud 
interior like L1004E \citep[e.g.,][]{Bergin2002,Tafalla2002,Lada2007}, while we assume
the linear $N$(C$^{18}$O)--$N$(H$_{2}$) conversion relation based on the results of \citet{Frerking1982}
whose measurements were done in much less dense regions. However, we shall use the simple
conversion relation in Equation (\ref{eq:NH2}) in this paper, because it is difficult at the moment
to assess precisely how much the depletion of C$^{18}$O is occurring in L1004E.



\clearpage


\begin{deluxetable}{lcrcccccccccc} 
\tabletypesize{\scriptsize}
\rotate
\tablewidth{0pc} 
\tablecaption{Mapping Observations  \label{tab:mapping_obs}} 
\tablehead{ 
 \colhead{}   &  \colhead{} &  \colhead{} &  \colhead{}  & \multicolumn{2}{c}{Receiver} &  \colhead{} & \multicolumn{3}{c}{Spectrometer} &  \colhead{}    &  \colhead{}   & \colhead{}  \\ 
  \cline{5-6}  \cline{8-10}  \\ 
 \colhead{Molecule} & \colhead{Transition}   & \colhead{Rest Frequency\tablenotemark{(1)}}  & \colhead{HPBW} &
  \colhead{Name} &  \colhead{$T_{\rm sys}$}  &  \colhead{} &  \colhead{Name} & \colhead{Band Width} & \colhead{$\Delta f_{\rm reso}$} &
  \colhead{$\Delta T_{\rm mb}$} & \colhead{$\Delta V_{\rm reso}$} & \colhead{Grid}  \\
 \colhead{} & \colhead{}   & \colhead{(GHz)}  & \colhead{(arcsec)} &
  \colhead{} &  \colhead{(K)}  & \colhead{} & \colhead{} &  \colhead{(MHz)} & \colhead{(kHz)} &  
  \colhead{(K)} & \colhead{(km s$^{-1}$)} & \colhead{(arcsec)}  \\ 
}
\startdata 
$^{12}$CO	&	$J=1-0$		&	115.271204 	&   15 	&  BEARS   &  340 & &   AC   &   32   &   38   &  1.25 & 0.22 & 10 \\
$^{13}$CO	&	$J=1-0$		&	110.201353 	&   16 	&  BEARS   &  320 & &   AC   &   32   &   38   &  0.94 & 0.14 & 10 \\
C$^{18}$O	&	$J=1-0$		&	109.782182 	&   16 	&  BEARS   &  310 & &   AC   &   32   &   38   &  0.53 & 0.14 & 10 \\
HC$_{3}$N	&	$J=5-4$		&	45.490319 	&   40 	&  S40         &  230 & &   AC   &   32   &   38   &  0.14 & 0.32 & 30 \\
CCS			&	$J_N=4_3-3_2$ &	45.379033 	&   40 	&  S40         &  230 & &   AC   &   32   &   38   &  0.15 & 0.32 & 30 \\
NH$_3$		&	$(J,K)=(1,1)$	&	23.694506 	&   75 	&  H22         &  140 & &   AC   &   16   &   19   &  0.08 & 0.34 & 30 \\
\enddata

\tablerefs{(1) \citet{Lovas1992}}

\end{deluxetable} 

\clearpage

\begin{deluxetable}{lcrccccccccc} 
\tabletypesize{\scriptsize}
\rotate
\tablewidth{0pc} 
\tablecaption{One Point Observations  \label{tab:point_obs}} 
\tablehead{ 
 \colhead{}   &  \colhead{} &  \colhead{} &  \colhead{}  & \multicolumn{2}{c}{Receiver} &  \colhead{} & \multicolumn{3}{c}{Spectrometer} &  \colhead{}    &  \colhead{}  \\ 
  \cline{5-6}  \cline{8-10}  \\ 
 \colhead{Molecule} & \colhead{Transition}   & \colhead{Rest Frequency\tablenotemark{(1)}}  & \colhead{HPBW} &
  \colhead{Name} &  \colhead{$T_{\rm sys}$}  &  \colhead{} &  \colhead{Name} & \colhead{Band Width} & \colhead{$\Delta f_{\rm reso}$} &
  \colhead{$\Delta T_{\rm mb}$} & \colhead{$\Delta V_{\rm reso}$}  \\
 \colhead{} & \colhead{}   & \colhead{(GHz)}  & \colhead{(arcsec)} &
  \colhead{} &  \colhead{(K)} & \colhead{} & \colhead{} &  \colhead{(MHz)} & \colhead{(kHz)} &  
  \colhead{(K)} & \colhead{(km s$^{-1}$)}  \\ 
}
\startdata 
CS			&	$J=2-1$		&	 97.980968 	& 18 & T100H   &  160  & &   AOS   &   40   &   37   &  0.09 &0.11 \\
C$^{34}$S	&	$J=2-1$		&	 96.412982 	& 18 & T100H   &  160 & &   AOS   &   40   &   37   &  0.25 & 0.12 \\
HCO$^{+}$	&	$J=1-0$		&	 89.188518 	& 19 & T100V   &  430 & &   AOS   &   40   &   37   &  0.04 & 0.12 \\
H$^{13}$CO$^{+}$	&	$J=1-0$	&	 86.754330 	& 19 & T100H   &  160 & &   AOS   &   40   &   37   &  0.04 &0.13 \\
NH$_3$		&	$(J,K)=(1,1)$	&	 24.139417 	& 75 &   H22         &  140 & &   AOS   &   40   &    37   &  0.03 &0.46 \\
NH$_3$		&	$(J,K)=(2,2)$	&	 23.870129 	& 75 &   H22         &  140 & &   AOS   &   40   &    37   &  0.03 &0.47 \\
NH$_3$		&	$(J,K)=(3,3)$	&	 23.722633 	& 75 &   H22         &  140 & &   AOS   &   40   &    37   &  0.02 & 0.47 \\
NH$_3$		&	$(J,K)=(4,4)$	&	 23.694506 	& 75 &   H22         &  140 & &   AOS   &   40   &    37   &  0.02  & 0.47 \\
\enddata

\tablerefs{(1) \citet{Lovas1992}}

\end{deluxetable} 

\clearpage

\begin{deluxetable}{ll} 
\tabletypesize{\scriptsize}
\tablewidth{0pc} 
\tablecaption{Physical Properties of L1004E  \label{tab:core}} 
\tablehead{ 
 \colhead{Quantities}  & \colhead{} \\ 
}
\startdata 
Mass & $1.1 \times 10^{4}$ M$_\odot$ \\
$T_ {\rm ex}$ & $8-12$ K \\
$\Delta V$(C$^{18}$O) & $1-3$ km s$^{-1}$ \\
$\tau$(C$^{18}$O) & $2-3$ \\
Size & $\sim 2 \times 5$ pc$^2$ \\
SFE & $\lesssim 1$ $\%$
\enddata

\end{deluxetable}

\begin{deluxetable}{cccccccrccc} 
\tabletypesize{\scriptsize}
\tablewidth{0pc} 
\tablecaption{IRAS Point Sources  \label{tab:iras}} 
\tablehead{ 
 \colhead{}  & \multicolumn{2}{c}{Coordinates} & \colhead{} & \multicolumn{4}{c}{Flux Density} &  \colhead{}    &  \colhead{}  &  \colhead{}  \\ 
  \cline{2-3}  \cline{5-8}  \\ 
 \colhead{IRAS No.} & \colhead{$\alpha_{\rm J2000}$}   & \colhead{$\delta_{\rm J2000}$}  & \colhead{} &
  \colhead{$F12$} &  \colhead{$F25$}  &  \colhead{$F60$} &  \colhead{$F100$} & 
  \colhead{Quality} & \colhead{C. C.} & \colhead{$L_{\rm IRAS}$} \\
 \colhead{} & \colhead{}   & \colhead{}  & \colhead{} &
  \colhead{(Jy)} &  \colhead{(Jy)} & \colhead{(Jy)} & \colhead{(Jy)} &  
  \colhead{} & \colhead{}  & \colhead{($L_\odot$)} \\ 
}
\startdata 
$21005+5217$ & $21^{\rm h}02^{\rm m}05.5^{\rm s}$ & $52{\arcdeg}28{\arcmin}54{\arcsec}$ & & 0.9932(6) & 2.580(5) & 7.521(9) & $<10.97 ($...$)$ & 3331 & AAAB & 21.8 \\
$21025+5221$ & $21^{\rm h}04^{\rm m}06.5^{\rm s}$ & $52{\arcdeg}33{\arcmin}47{\arcsec}$ & & 0.3691(11) & 1.907(28) & 8.101(10) & 39.96(17) & 3333 & CAAA & 35.9 \\
\enddata

\tablecomments{Numbers in the parentheses in the columns for $F12-F100$ are the uncertainties in units of percent. $L_{\rm IRAS}$ is the bolometric luminosity 
integrated over the four IRAS bands by the method of \citet{Myers1987}.}

\end{deluxetable} 

\begin{deluxetable}{ccccccccc} 
\tabletypesize{\scriptsize}
\setlength{\tabcolsep}{0.04in}
\tablewidth{0pc} 
\tablecaption{ Counterparts of the IRAS Sources \label{tab:counterparts}} 
\tablehead{ 
 \colhead{}  & \multicolumn{4}{c}{WISE} & \colhead{} & \multicolumn{3}{c}{2MASS}  \\ 
  \cline{2-5}  \cline{7-9}  \\ 
 \colhead{IRAS No.} & \colhead{W1}  & \colhead{W2} &  \colhead{W3} &  \colhead{W4}  &  \colhead{} &  \colhead{$J$} & \colhead{$H$} & \colhead{$K_{\rm S}$} \\
\colhead{} & \colhead{(mag)}  & \colhead{(mag)} &  \colhead{(mag)} &  \colhead{(mag)}  &   \colhead{} &  \colhead{(mag)} & \colhead{(mag)} & \colhead{(mag)} \\
}
\startdata 
$21005+5217$ &$9.165\pm0.021$ & $7.207\pm 0.020$ & $3.601\pm0.015$ & $1.095\pm0.010$ & & $13.457\pm0.024$ & $11.701\pm0.021$ & $10.519\pm0.015$ \\
$21025+5221$ &$10.160\pm0.025$ & $7.621\pm 0.020$ & $5.297\pm0.014$ & $2.039\pm0.015$ & & $>18.296$ & $>15.779$ & $13.846\pm0.074$ \\
\enddata
\tablecomments{Counterparts of IRAS 21005+5217 and 21025+5221 are J210205.43+522854.2 and J210407.18+523350.0 in the WISE PSC, and 717809903 and 717897656 in the 2MASS PSC, respectively.}

\end{deluxetable} 

\begin{deluxetable}{lcccc} 
\tabletypesize{\scriptsize}
\tablewidth{0pc} 
\tablecaption{Parameters of YSOs  \label{tab:modelpara}} 
\tablehead{
 \colhead{IRAS No.}  & \colhead{$A_V$}   &  Age               & $M_{\rm star}$ & $L_{\rm total}$ \\
 \colhead{}                     & \colhead{(mag)} & \colhead{(yr)}& \colhead{($M_{\odot}$)} & \colhead{($L_{\odot}$)}  \\
}
\startdata 
$21005+5217$ & 13.68   & $1.61\times 10^{6}$ & 3.63 & $2.06\times 10^2$  \\
$21025+5221$ & 27.06   & $5.99 \times 10^{4}$ & 2.44 & $5.35 \times 10^1$  \\
\enddata
\tablecomments{The values are the best fitting parameters for a model developed by \citet{Robitaille2007}.}

\end{deluxetable} 

\begin{deluxetable}{cccccccccccccc} 
\tabletypesize{\scriptsize}
\rotate
\setlength{\tabcolsep}{0.05in}
\tablewidth{0pc} 
\tablecaption{ Properties of Filaments and Subcores \label{tab:filaments}} 
\tablehead{
\colhead{No.}  & \colhead{$V_{\rm range}$} & \colhead{$\alpha_{\rm J2000}$} &\colhead{$\delta_{\rm J2000}$}  & \colhead{$N$(H$_2$)} &  \colhead{$ \Delta V$} &\colhead{$M_{\rm LTE}$} &  \colhead{$ \Delta {\overline V}$}  & \colhead{$S$} & \colhead{$R$}& \colhead{$M_{\rm vir}$} & \colhead{Ellipticiticy} &  \colhead{$M_{\rm cr}$} & \colhead{$M_{\rm LTE}/L$}\\
\colhead{}  & \colhead{(km s${^{-1}}$)}  & \colhead{} &\colhead{}  & \colhead{(10$^{22}$ cm$^{-2}$)} & \colhead{(km s${^{-1}}$)} & \colhead{($M_{\odot}$)}  & \colhead{(km s${^{-1}}$)}  & \colhead{(pc$^2$)} & \colhead{(pc)}  & \colhead{($M_{\odot}$)}  & \colhead{$\alpha$} & \colhead{($M_{\odot}$ pc$^{-1}$)} & \colhead{($M_{\odot}$ pc$^{-1}$)}  \\
}
\startdata 
 1&$(-3.5,-0.3)$&$21^{\rm h}04^{\rm m}03.2^{\rm s}$ & $52{\arcdeg}34{\arcmin}06{\arcsec}$ & 3.1&1.58& 170&1.70& 0.34&0.33& 199&1.00 & 1158 & 259\\
 2&$(-4.0,-2.0)$&$21^{\rm h}03^{\rm m}57.7^{\rm s}$ & $52{\arcdeg}28{\arcmin}36{\arcsec}$ & 2.1&0.44&230&1.52& 0.52&0.41& 196&0.91 &  90& 277\\
 3&$(-2.2,-0.9)$&$21^{\rm h}03^{\rm m}55.5^{\rm s}$ & $52{\arcdeg}36{\arcmin}46{\arcsec}$ & 1.9&0.63&177&1.08&0.45&0.38&  ~91&0.17 & 186& 96\\
 4&$(-2.2,-0.9)$&$21^{\rm h}03^{\rm m}38.0^{\rm s}$ & $52{\arcdeg}36{\arcmin}16{\arcsec}$ & 2.1&0.77&326&1.04&0.64&0.45& 102&0.37 & 277& 221\\
 5&$(-2.2,-0.9)$&$21^{\rm h}03^{\rm m}34.7^{\rm s}$ & $52{\arcdeg}32{\arcmin}16{\arcsec}$ & 2.1&1.20&420&1.39&0.76&0.49& 200&0.20 & 668& 192 \\
 6&$(-4.4,-2.5)$&$21^{\rm h}03^{\rm m}16.1^{\rm s}$ & $52{\arcdeg}31{\arcmin}36{\arcsec}$ & 1.9&0.73&590&1.88&1.30&0.64& 478&0.42 & 249& 295 \\
 7&$(-3.0,-0.3)$&$21^{\rm h}02^{\rm m}22.5^{\rm s}$ & $52{\arcdeg}28{\arcmin}24{\arcsec}$ & 4.5&1.78&793&1.78&1.04&0.57& 380&0.50 & 1471& 488 \\
 8&$(-4.6,-1.0)$&$21^{\rm h}02^{\rm m}20.7^{\rm s}$ & $52{\arcdeg}18{\arcmin}44{\arcsec}$ & 3.0&1.48&212&1.89&0.22&0.27& 197&0.63 & 1021& 314 \\
 9&$(-3.6,~~0.0)$&$21^{\rm h}02^{\rm m}00.0^{\rm s}$ & $52{\arcdeg}19{\arcmin}42{\arcsec}$ & 3.0&2.54&165&2.94&0.32&0.32& 579&0.77 & 2994& 223 \\
10&$(-1.7,~~0.0)$&$21^{\rm h}01^{\rm m}42.4^{\rm s}$ & $52{\arcdeg}22{\arcmin}31{\arcsec}$ & 1.9&0.42&107&1.26&0.35&0.33& 110&0.43 & 82& 106 \\
11&$(-3.6,-1.7)$&$21^{\rm h}01^{\rm m}34.7^{\rm s}$ & $52{\arcdeg}22{\arcmin}30{\arcsec}$ & 2.1&1.09&415&2.42&0.95&0.55& 670&0.48 & 551& 261 \\
12&$(-3.6,-1.8)$&$21^{\rm h}01^{\rm m}25.7^{\rm s}$ & $52{\arcdeg}27{\arcmin}50{\arcsec}$ & 2.1&1.45&491&1.51&0.81&0.51& 240&0.28 & 983& 255 \\
\enddata
\tablecomments{
Rather round filaments (Nos. 1, 2, 8, and 9) with a high ellipticity of $\alpha>0.5$ are called ``subcores" in Section \ref{subset:filaments}.
Values in the table are measured based on the C$^{18}$O data
(see Section \ref{subsubsect:stability}). Some filaments
and subcores (Nos. 2, 3, 5, 6, 9, 10, and 11) 
are partially contaminated by another filament at different velocities, and we derived  $ \Delta V$ (the FWHM line width at the peak positions)
and $ \Delta {\overline V}$ (the line width averaged over the surface $S$) by applying Gaussian fitting with 2 components.
}
\end{deluxetable}

\clearpage


\clearpage

\clearpage

\clearpage

\clearpage

\clearpage

\clearpage

\clearpage

\clearpage

\clearpage

\clearpage

\clearpage

\clearpage

\clearpage

\clearpage

\clearpage

\clearpage

\clearpage

\end{document}